\documentclass[reprint, amsmath, amssymb, aps, prl]{revtex4-1}

\usepackage{amssymb}
\usepackage{graphicx}
\usepackage{dcolumn}
\usepackage{bm}
\usepackage{upgreek}
\usepackage{epstopdf}
\usepackage{extarrows}
\usepackage{amssymb}
\usepackage{amsmath}
\usepackage{color}
\begin{document}

\preprint{APS/123-QED}

\title{Imaginary Gauge Field and Non-Hermitian Topological Transition Emerging Through Attenuation-Gauge Duality in Conservative Systems}

\author{Haoran Nie$^{1}$, Chaoran Jiang$^{1}$, Xiangying Shen$^{2*}$, Lei Xu$^{1, 3*}$}

\affiliation{$^1$ Department of Physics, The Chinese University of Hong Kong, Hong Kong, China
\\$^2$School of Science, Shenzhen Campus of Sun Yat-sen University, Shenzhen, China\\$^3$Shenzhen Research Institute, The Chinese University of Hong Kong, Shenzhen 518057, China}

\date{\today}

\begin{abstract}
Non-Hermitian physics traditionally relies on active gain–loss modulation or non-reciprocal couplings, which often introduce significant complexity, compromise stability, and offer very limited scalability in conservative systems. Here we propose an attenuation-gauge duality paradigm in which non-Hermitian topology emerges within fully passive, conservative systems through coupling to a structured reservoir. We derive that a spatially varying reservoir can establish an attenuation-gauge duality, where the spatial variation manifests as an emergent imaginary gauge field in the effective dynamics. It drives the boundary accumulation of skin modes while preserving energy conservation, analogous to Feshbach projection in quantum open systems. We validate this universal wave paradigm via macroscopic mechanical metamaterials, demonstrating that the direction of the skin effect can be reversed by tuning a single passive coupling parameter $t_\perp$, driven by a topological phase transition characterized by the spectral winding number. This framework also allows for a nonlinear extension, where amplitude-dependent coupling can induce intrinsic topological transitions.
\end{abstract}

\maketitle

Non-Hermitian physics, originally developed to describe open quantum systems, has expanded the framework of wave control by introducing specific features absent in Hermitian counterparts \cite{shenTopologicalBandTheory2018a,yaoNonHermitianChernBands2018,jiangTopologicalInvariantsPhase2018}. Central to these developments are exceptional points (EPs) and the non-Hermitian skin effect (NHSE), which enable extreme spectral sensitivity and boundary-selective localization \cite{miriLargeAreaSinglemode2012,fuModeControlQuasiPT2020,hodaeiParitytimeSymmetricMicroring2014,zhangEdgeTheoryNonHermitian2024,zhangUniversalNonHermitianSkin2022,wangTopologicalComplexenergyBraiding2021,kawabataHigherorderNonHermitianSkin2020,okumaTopologicalOriginNonHermitian2020a,yokomizoNonBlochBandTheory2019,imuraGeneralizedBulkedgeCorrespondence2019,borgniaNonHermitianBoundaryModes2020}. These effects have stimulated implementations across photonic, acoustic, and mechanical platforms\cite{zhaoNonHermitianTopologicalLight2019,zhangAcousticNonHermitianSkin2021a,gaoNonHermitianRouteHigherorder2021,fanReconfigurableTopologicalModes2023,gaoControllingAcousticNonHermitian2024} and are being extended to nonlinear settings where intensity can drive topological changes \cite{liTopologicalSwitchNonHermitian2020,yangNonHermitianHopflinkExceptional2019,longNonHermitianTopologicalSystems2022,liangProbingTopologicalPhase2023a,zhangBulkboundaryCorrespondenceNonHermitian2020,guoTheoreticalPredictionNonHermitian2022,soneTransitionTopologicalChaotic2025,kawabataHopfBifurcationNonlinear2025,xiaNonlinearTuningPT2021a,daiNonHermitianTopologicalPhase2023,okuma2020topological,zhang2020correspondence}. However, most experimental implementations of the non-Hermitian physics rely on engineered gain-loss distributions or active non-reciprocal couplings \cite{zhouNonHermitianTopologicalMetamaterials2020a,scheibnerNonHermitianBandTopology2020,ghatakObservationNonHermitianTopology2020,mengNonHermitianTopologicalCoupler2021}. These approaches often demand complex feedback circuits, precise energy injection, or odd-elasticity interactions \cite{veenstraNonreciprocalTopologicalSolitons2024,scheibnerOddElasticity2020b,wangNonHermitianMorphingTopological2022,wangNonHermitianTopologyStatic2023}, needing energy to maintain and introducing significant complexity that can undermine stability and scalability. Thus, a key open question remains: can such phenomena be realized without relying on active or explicitly non-reciprocal elements? Achieving controllable non-Hermitian topology in a fully passive and conservative setting would not only simplify realistic architectures but also bridge the gap between non-Hermitian theory and robust engineering applications. These include, but are not limited to, vibration damping, energy harvesting, and phononic information processing \cite{zhou2016research,dykstra2023buckling,wang2016tunable,liao2021acoustic,ryu2019hybrid,cao2023observation,chen2021topological}.

In this work, we establish a general paradigm for emergent non-Hermiticity in passive conservative systems via attenuation-gauge duality. Instead of introducing active elements, we partition the total closed system into a primary subsystem and a structured environment (reservoir), coupled through spatially varying interactions. By employing the Feshbach projection, we derive from first principles that the interaction with the environment manifests as a frequency-dependent self-energy term in the subsystem's effective Hamiltonian. Crucially, we uncover an attenuation-gauge duality in the adiabatic limit, where a spatially varying reservoir rigorously maps to an emergent imaginary gauge potential. Furthermore, we demonstrate that the localized skin modes in our passive system are adiabatically connected to the non-Hermitian skin effect, topologically protected by the non-closing of the point gap. This geometric mechanism naturally drives the non-Bloch accumulation of skin modes from purely conservative dynamics. To experimentally demonstrate that non-Hermitian phenomena universally extend beyond quantum systems, we validate this paradigm using a mechanical oscillator network. We show that the direction of the skin effect can be reversed by tuning a single passive parameter, the interchain coupling stiffness $t_\perp$, and identify this reversal as a topological phase transition characterized by a change in the spectral winding number. The same framework also provides a minimal route to passive nonlinear extensions, where amplitude-dependent coupling can trigger topological switching analogous to the frontier research of Kerr effect in nonlinear optics.

\begin{figure*}[htpb]
\begin{center}
\centerline{\includegraphics[width=0.85\linewidth]{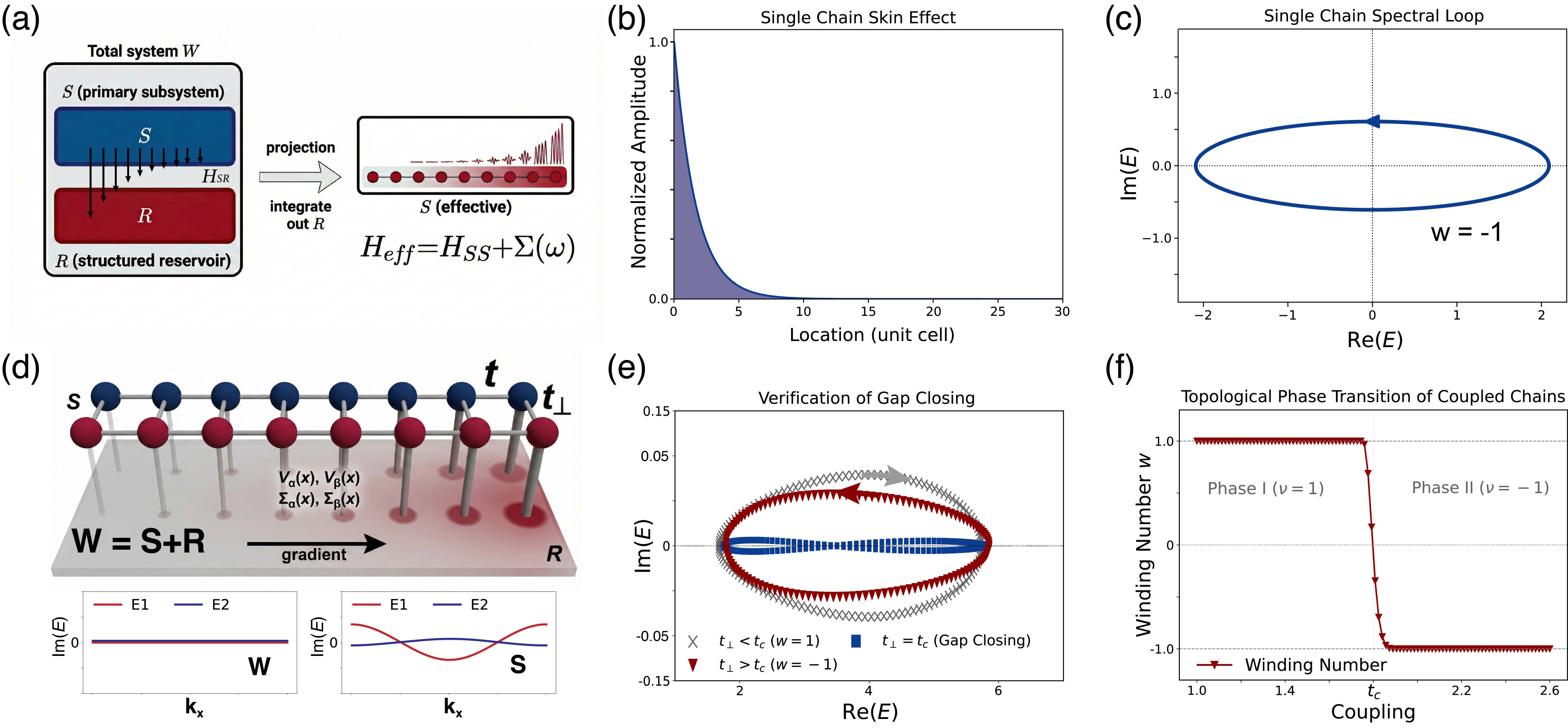}}
\caption{\label{fig.1} Emergence of non-Hermitian dynamics in a passive conservative system via structured reservoir engineering. (a) Theoretical paradigm: Partitioning the total Hermitian system $W$ into a subsystem $S$ and a structured reservoir $R$. A spatial varying interaction $v(x)$ induces a quadratic decay profile $\Gamma(x) \propto |v(x)|^2$, establishing the Attenuation-Gauge duality.
(b,c) Numerical validation on a single chain ($N=30$, $\bar{\lambda} \approx 0.3$). 
The projection-induced decay generates a complex spectrum of subsystem $S$, resulting in the exponential skin mode envelope in (b) and the corresponding spectral winding loop in (c).
(d) A coupled double-chain model representing the subsystem $S$, characterized by intra-chain coupling $t$ and inter-chain coupling $t_\perp$. The structured reservoir induces spatial varying outflow $\Gamma_\alpha(x)$ and $\Gamma_\beta(x)$ on the respective chains, where their opposite gradients compete. The imaginary part of $W$ and $S$'s energy spectra are exhibited at the bottom panels. Clearly $W$ is conserved with 0 imaginary part, while $S$ does have non-zero imaginary part. (e) Verification of the topological phase transition. The low-energy spectral trajectory of the determinant crosses at the critical coupling $t_\perp = t_c$ (navy line), confirming the point-gap closing mechanism ($\det H_{\text{ref}}=0$) that separates the $w=1$ (grey dashed) and $w=-1$ (red solid) topological phases. (f) Winding number jumps from 1 to -1 at the critical coupling $t_\perp=t_c$. }
\end{center}
\vspace{-0.5cm}
\end{figure*}

\noindent\textit{---System partition and attenuation-gauge duality.} To rigorously demonstrate the emergence of non-Hermitian topology in a passive setting, we partition the total system $W$ into a primary subsystem $S$ and a structured reservoir $R$, as illustrated in Fig.~\ref{fig.1}(a). The dynamics of the total system is governed by a Hermitian Hamiltonian $H_{\text{tot}}$, reflecting strict energy conservation. In the frequency domain, $H_{\text{tot}}$ takes the block form:
\begin{equation}
\label{eq:Htot}
H_{\text{tot}} =
\begin{pmatrix}
H_{SS} & H_{SR} \\
H_{RS} & H_{RR}
\end{pmatrix},
\end{equation}
where $H_{SS}$ and $H_{RR}$ describe the internal dynamics of the subsystem and the reservoir, respectively, and $H_{SR} = H_{RS}^\dagger$ denotes the coupling between them. By integrating out the reservoir degrees of freedom $\psi_R$ via Feshbach projection (see Supplemental Material Sec.~I.A), we obtain an effective equation of motion for the subsystem $H_{\text{eff}}(\omega)\psi_S = [H_{SS} + \Sigma(\omega)]\psi_S = \omega^2\psi_S$, governed by the effective Hamiltonian:
\begin{equation}
\label{eq:Heff}
H_{\text{eff}}(\omega) = H_{SS} + \Sigma(\omega), \quad \Sigma(\omega) = H_{SR} G_R(\omega) H_{RS}.
\end{equation}
Here, $\Sigma(\omega)$ is the frequency-dependent self-energy operator, and $G_R(\omega)=(\omega I-H_{RR})^{-1}$ is the Green's function of the eliminated reservoir degrees of freedom. In the reduced description of the subsystem $S$, the outflow of energy into the reservoir manifests effectively as an anti-Hermitian contribution. Consequently, the passivity of the total system $W$ imposes the constraint $\mathrm{Im}\,[\Sigma(\omega)]\le 0$, ensuring that the effective dynamics exhibit attenuation. We emphasize that this attenuation describes the projection-induced amplitude reduction within $S$, which is not caused by actual damping.

Next, we combine the previous paradigm with a spatial structuring of the system-reservoir coupling. Specifically, we suppose the subsystem-reservoir coupling strength $v(x)$ follows a \textit{monotonic} spatial variation, for example $v(x) \propto e^{\eta x}$, as conceptually illustrated in Fig.~\ref{fig.1}(a). Under the local-coupling and wide-band approximations (see Supplemental Material Secs.~I.B), this spatially varying coupling translates into a spatially varying imaginary potential in the self-energy, $\Sigma(x) \approx \Delta(x) - i\Gamma(x)$, where the outflow-induced attenuation rate $\Gamma(x)$ scales as $\Gamma(x) \propto |v(x)|^2 \propto e^{2\eta x}$. Then, this spatial inhomogeneity can be correlated with complex momentum using an attenuation–gauge duality framework. In the adiabatic limit ($\eta a \ll 1$) and the weak non-Hermiticity regime ($|\kappa a| \ll 1$, with $\kappa$ explained later), the spatial variation of the decay rate $\Gamma(x)$ is slow compared to the lattice spacing. By employing a Wentzel-Kramers-Brillouin (WKB) analysis for the stationary eigenmodes $\psi(x) \sim \exp(i\int k(x)dx)$, with a position-dependent complex wavevector $k(x) = k_R(x) + i\kappa(x)$, we find that the local imaginary momentum $\kappa(x)$ is directly determined by the attenuation rate $\Gamma(x)$ (see Supplemental Material Sec.~I.C):
\begin{equation}
\label{eq:WKB}
\kappa(x) \approx \frac{\Gamma(x)}{v_g(k_R)},
\end{equation}
where $v_g$ is the group velocity. This relation reveals the physical essence of the duality: the spatial attenuation rate $\Gamma(x)$ in real space is dynamically equivalent to a local imaginary shift $\kappa(x)$ in the momentum space, mathematically corresponding to a gauge transformation.

This local duality allows us to define a global topological invariant through a coarse-graining procedure. By averaging the accumulated decay over the system length $L$, we define an emergent imaginary gauge field $\bar{\lambda}$:
\begin{equation}
\label{eq:WKB_ratio}
\bar{\lambda} \equiv \frac{1}{L}\int_0^L \kappa(x)\,dx \approx \frac{1}{L v_g}\int_0^L \Gamma(x)\,dx.
\end{equation}
This emergent gauge field $\bar{\lambda}$ characterizes the net non-reciprocal accumulation of the bulk modes, $|\psi(L)|/|\psi(0)| \sim e^{-\bar{\lambda}L}$. Consequently, we can construct a translationally invariant topological representative, the reference Hamiltonian $H_{\text{ref}}(k)$, by applying the imaginary gauge transformation $k \to k - i\bar{\lambda}$ to its original Hermitian counterpart $H_{\text{SS}}$: $H_{\text{ref}}(k) \equiv H_{\text{SS}}(k - i\bar{\lambda})$. This naturally induces an imaginary component and non-Hermitian effects.

\noindent\textit{---Chain model with spectral winding topology.} For a tight-binding chain, this transformation into $H_{\text{ref}}(k)$ naturally yields the non-reciprocal Hatano-Nelson model. The topological phase can be classified by the spectral winding number: $w = \frac{1}{2\pi i} \oint_{-\pi}^{\pi} \frac{d}{dk} \ln \det [H_{\text{ref}}(k) - E_{\text{C}}] \, dk$. Here, the passive structural gradient $\Gamma(x)$ plays the same role as an active non-reciprocal gauge field. The gauge field induces non-Hermitian skin mode in a single chain system, as depicted in Fig.~\ref{fig.1}(b). The localization of mode is governed by the winding direction of $H_{\text{ref}}$ in the complex plane. Specifically, the winding direction corresponds to $w=-1$ (counter-clockwise, see Fig.~\ref{fig.1}(c)) or $w=1$ (clockwise), which determines whether the skin modes localize at the left or right edge.

To harness this mechanism for topological control, we construct the subsystem $S$ as a coupled double-chain lattice, as shown in Fig.~1(d). This general tight-binding model consists of two chains ($\alpha$ and $\beta$) with intra-chain coupling $t$ and inter-chain coupling $t_\perp$. Moreover, the two chains couple to the reservoir with opposing spatial decay rates, $\Gamma_\alpha(x)$ and $\Gamma_\beta(x)$, as shown in Fig.~1(d). In our Attenuation-Gauge Duality framework, these opposing spatial decays manifest as competing imaginary gauge fields $\bar{\lambda}_\alpha$ and $\bar{\lambda}_\beta$ in the effective $2\times 2$ reference Hamiltonian $H_{\text{ref}}^{2\times 2}(k)$. Crucially, as we tune the inter-chain coupling $t_\perp$ from low to high, there exists a critical value $t_c$, at which the gap of the determinant $\det[H_{\text{ref}}^{2\times 2}(k)]$ closes and its winding direction reverses, as demonstrated in Fig.~\ref{fig.1}(e) (see details in Supplemental Material Sec.~I.E). This confirms the point-gap closing condition required for the topological phase transition.

Physically, the topological invariant $w$ corresponds to the winding number of the lower energy band $E_-(k)$ (fundamental modes). Since the winding number of the determinant is topologically equivalent to that of $E_-(k)$ in the gap-open regime, the gap-closing and band-crossing at $t_\perp = t_c$ observed in Fig.~\ref{fig.1}(e) confirms the inversion of the spectral winding direction, thereby producing a topological phase transition. In Fig.~\ref{fig.1}(f), we present the value of $w$ as a function of $t_\perp$, where a topological phase transition is clearly identified by a sharp change at $t_\perp = t_c$. This phase transition corresponds to a reversal of the skin mode localization from one boundary to the other, as demonstrated later.

\begin{figure*}[htpb]
\begin{center}
\centerline{\includegraphics[width=0.85\linewidth]{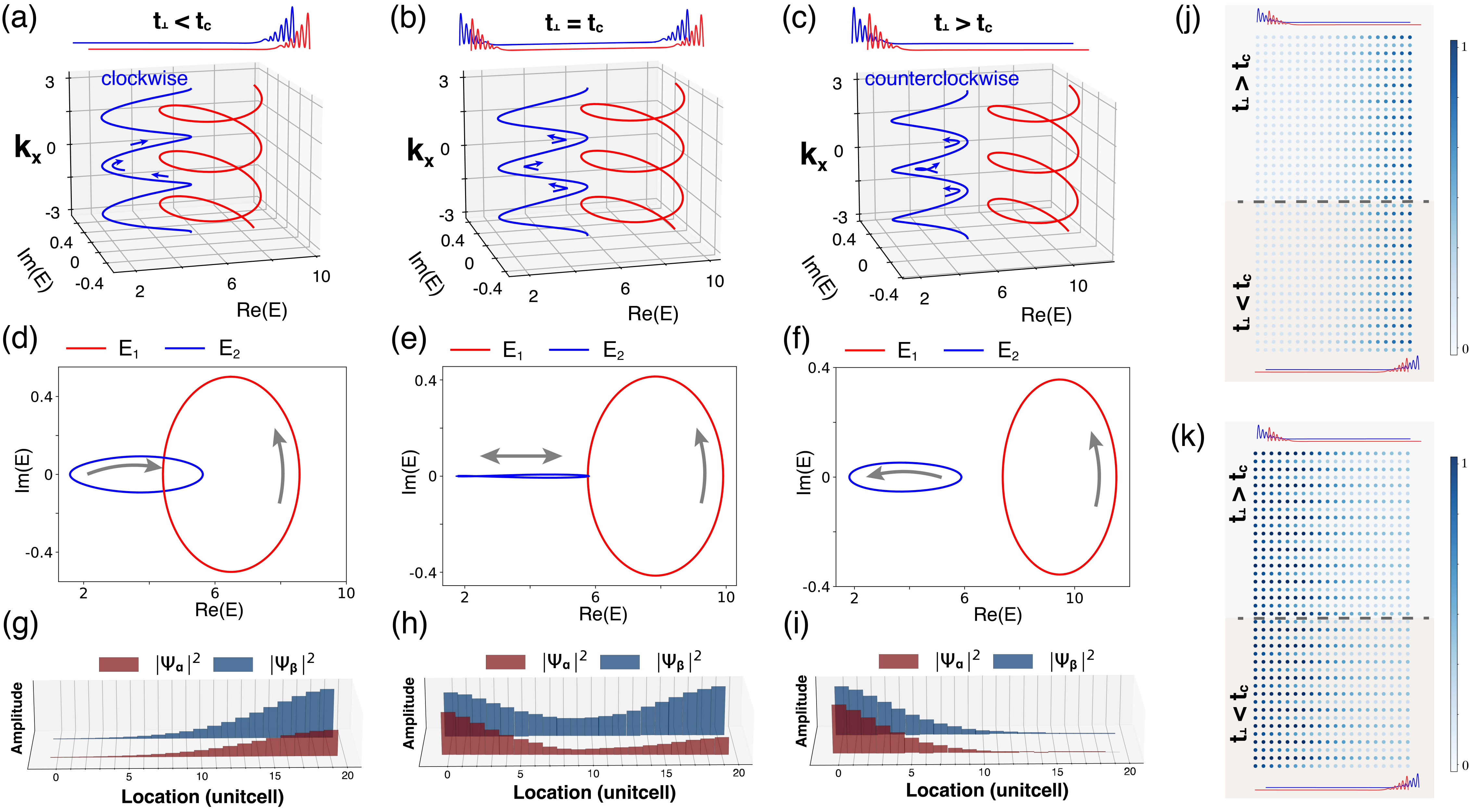}}
\caption{\label{fig.2}Topological phase transition and skin mode reversal driven by passive coupling. (a)--(c) Evolution of the skin mode localization as the inter-chain coupling $t_\perp$ increases across the critical point $t_c$. The mode switches from the right boundary ($t_\perp < t_c$) to the left boundary ($t_\perp > t_c$), governed by the point-gap closing and the resulting inversion of the spectral winding between the two coupled chains with opposing gradient in grounded springs. (d)--(f) A 2D view of the complex energy spectrum $E(k)$ shows that the spectral winding direction of the lower energy band (indicated in blue) reverses as $t_\perp$ crosses $t_c$, resulting in a change in the winding number from $w=1$ to $w=-1$. (g)--(i) Corresponding vibration amplitude profiles of the mechanical chains, confirming the spatial reversal of the skin modes, where $t=1,\nabla v_\alpha=0.2,\nabla v_\beta=-0.05, t_c\simeq2.1$ and $v_{\alpha}=2, v_{\beta}=1,$ at left end point. (j, k) Collective skin effect in a 2D stacked configuration. Despite the opposing gradients in the top and bottom halves, the bulk modes collectively localize to a single boundary, dictated by the dominant gradient direction.}
\end{center}
\vspace{-0.5cm}
\end{figure*}

\noindent\textit{---Non-Hermitian phase transition in coupled mechanical chains.} To physically verify this paradigm, we construct a macroscopic mechanical metamaterial composed of two coupled chains, $\alpha$ and $\beta$, the same as the numerical model illustrated in Fig.~\ref{fig.1}(d). As detailed in Supplemental Material Sec.~I.F, a non-zero $\mathrm{Im}[\Sigma]$ arises even when coupling to a lossless reservoir (e.g., a semi-infinite spring chain) that supports energy outflowing modes. The system dynamical matrix can be mapped to a tight-binding model Hamiltonian where the spring stiffness dictates the hopping parameters \cite{nie2025designing} (see Supplemental Material Sec.~II). Crucially, the structured reservoir can be implemented using grounded vertical springs that connect each mass to a rigid base. By spatially varying the stiffness of these grounded springs, we engineer the requisite monotonic spatial profiles $v_\alpha(x)$ and $v_\beta(x)$.

Although the total mechanical system is conservative (lossless), the vibrational eigenmodes are physically distributed between the double chains (primary subsystem $S$) and the grounded springs (reservoir $R$). The spatial variation in the grounded stiffness $v(x)$ modulates the potential distribution: in regions of higher stiffness, a larger fraction of the local potential energy is stored in the reservoir $R$. Consequently, for a propagating mode, the vibrational amplitude projected onto the subsystem $S$ naturally decreases as it moves into the region of stiffer grounded springs. Importantly, the total mechanical energy remains conserved; what attenuates is only the component projected onto the primary subsystem $S$. This projection-induced spatial localization yields an envelope $|\psi_S(x)| \sim e^{-\int \kappa dx}$, which aligns with the theoretical framework of Attenuation-Gauge Duality. When the mechanism transitions from dissipation-driven ($\kappa \propto |v|^2$) to potential-driven ($\kappa \propto |v|$) in the conservative limit, the non-Hermitian topological nature is robustly preserved by the coupling gradient $\nabla |v(x)|$, demonstrating the adiabatic continuity from the conservative limit to the standard non-Hermitian regime (see Supplemental Material Sec.~I.F). Note that the non-Hermitian skin mode exists as long as $v(x)$ is monotonic, which can be as simple as a linear variation.

The topological phase of this system is characterized by the spectral winding number $w$, which is defined through the reference Hamiltonian $H_{\text{ref}}$. Our system undergoes a topological phase transition driven solely by the passive inter-chain coupling $t_\perp$, without the need for active gain-and-loss control typically required in non-Hermitian systems. When the inter-chain coupling increases from $t_\perp < t_c$ to $t_\perp > t_c$, the winding direction of the lower energy band reverses, the winding number $w$ changes from -1 to 1, and the localization of eigenmodes shifts from one boundary to the other, as shown in Fig.\ref{fig.2}(a-c). These specific results in mechanical double chains are consistent with the general theoretical framework in Fig.~\ref{fig.1}.

The change in winding direction is more clearly visualized in the 2D plots of Fig.~\ref{fig.2}(d-f): the lower-energy blue band’s winding direction transitions from clockwise to counterclockwise, indicating a distinct topological phase transition. It is important to note that only the lower energy band (blue) is crucial for this transition. Mathematically, the total winding number is the sum of the winding numbers of all energy bands. However, the phase transition point (gap closing) occurs only when the lower energy band $E_2$ crosses the origin, while the higher energy band $E_1$ remains far from the origin during this process, contributing nothing to the topological change. Thus, monitoring the winding behavior of the lower energy band is entirely equivalent to tracking the topological phase transition of the entire system. In Fig.~\ref{fig.2}(g-i), we show the calculated vibration amplitude distributions, which clearly demonstrate the presence of skin modes localized at the boundaries. Moreover, after $t_\perp$ crosses the critical value $t_c$, the skin modes reverse their localization, confirming the topological phase transition.

This attenuation-gauge duality is universal and also applies to 2D situation as depicted in Fig.~\ref{fig.2}(j,k), where identical double chains are stacked. The top and bottom halves belong to distinct regions: $t_\perp > t_c$ in the top and $t_\perp < t_c$ in the bottom. Surprisingly, skin modes are not distributed separately at the left and right edges according to the upper and lower halves in the entire 2D system; Instead overall modes localize to a single boundary, determined by the dominant player in the competition between the two regions and winding number of the ground state, as shown in Fig.~\ref{fig.2}(j,k) (see details in Supplemental Material Sec.~III). This collective behavior provides deeper insights into the attenuation-gauge duality. In 2D system, the interaction between the two regions hybridizes their distinct modes. The global topology of the system cannot be reduced to a simple superposition of its constituent parts; rather, it is governed by the dominant effective decay within the reservoir, which minimizes the imaginary energy penalty. As characterized in Supplemental Material Sec.~IV, the ``stronger" gradient ultimately dictates the winding number for the entire ground state manifold. This finding confirms that the emergent non-Hermiticity generates a robust, global topological phase that not only manifests in 1D but also extends to higher-dimensional complex assemblies.

\begin{figure*}[htpb]
\begin{center}
\centerline{\includegraphics[width=0.7\linewidth]{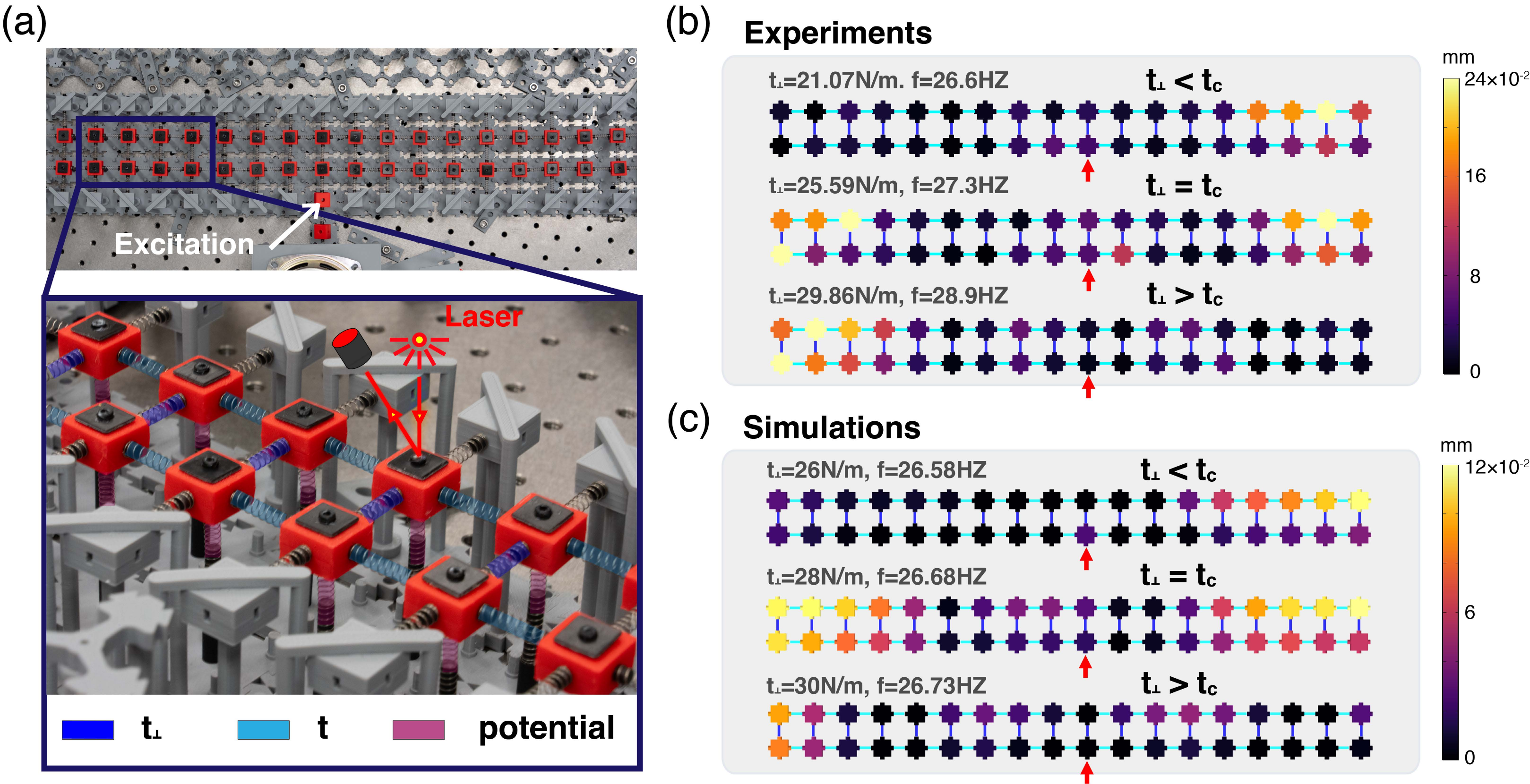}}
\caption{\label{fig.3}\textbf{Experimental validation of the passive skin mode reversal.} (a) Photograph of the mechanical double-chain prototype (top) and detailed structure (bottom). The inter-chain coupling $t_\perp$ is continuously tunable via movable clamps on the connecting springs, while the onsite potential gradients are engineered by the grounded springs. (b) Experimental measurements and (c) numerical simulations of the steady-state vibrational modes at 26.57{Hz}. As $t_\perp$ is tuned from 21.07{N/m} ($<t_c$) to 29.86{N/m} ($>t_c$), the localization of vibrational energy sharply switches from the right to the left boundary. This observation quantitatively confirms the theoretically predicted topological phase transition induced by passive structured coupling.}
\end{center}
\vspace{-0.5cm}
\end{figure*}

\noindent\textit{---Coupled chains paradigm in mechanical metamaterials.} To experimentally validate the proposed paradigm, we construct a macroscopic mechanical metamaterial consisting of two coupled oscillator chains, as depicted in Fig.~\ref{fig.3}(a). The structured reservoir is physically implemented using grounded vertical springs with stiffness varying linearly with location, which establishes the requisite projected outflow potentials $\Gamma_\alpha(x)$ and $\Gamma_\beta(x)$ in our primary subsystem description. A key feature of our setup is the precise tunability of the inter-chain coupling $t_\perp$. By adjusting the effective length of the horizontal connecting springs, $t_\perp$ can be continuously tuned across the topological critical point $t_c$, while keeping all other parameters unchanged. We excite the system at its ground eigenfrequency and measure the steady-state vibrational amplitude using a high-precision laser sensor. The experimental observations, presented in Fig.\ref{fig.3}(b), show excellent agreement with the numerical simulations in Fig.\ref{fig.3}(c). In the weak coupling regime ($t_\perp < t_c$), the vibrational energy is strongly localized at the right boundary, consistent with the topological winding number $w=1$, dictated by the dominant projected outflow gradient on the $\alpha$-chain. As $t_\perp$ is tuned beyond the critical threshold ($t_\perp > t_c$), a sharp reversal of the skin mode to the left boundary is observed, corresponding to $w=-1$. This clear and acute switching provides direct experimental confirmation that the passive interplay of the attenuation-gauge duality drives a topological phase transition corresponds to the change of winding number, verifying the theoretical paradigm.

\noindent\textit{---Nonlinear extension.} This framework can naturally extend beyond the linear regime to the nonlinear regime, which represents a frontier in non-Hermitian research. For instance, the linear environment-subsystem coupling of springs can be replaced with nonlinear elements by introducing an amplitude-dependent interaction: $v(x,|\psi|) \sim g|\psi|^2$, where the coupling $v$ depends nonlinearly on the mode amplitude $|\psi|$ with a coupling strength $g$. Such couplings can be implemented using passive materials, such as magnets (see Supplemental Material for details). These nonlinear couplings produce a response analogous to the Kerr effect, which can be further customized through structural design. With this nonlinear interaction, the non-Hermitian skin modes can transition from localizing at the right boundary to the left boundary simply by increasing the nonlinearity strength $g$, while keeping all other conditions unchanged. Simultaneously, the winding number $w$ of the low-energy band shifts from $w=1$ to $w=-1$, signaling a topological phase transition similar to the one induced by tuning the inter-chain coupling $t_\perp$ shown previously. This demonstrates a nonlinearity-induced phase transition in both the band topology and the non-Hermitian skin modes (see Supplemental Material Sec.~IV). The nonlinear result underscores the potential for intrinsic, excitation-driven topological switching in a fully passive platform, opening new degrees of freedom for exploring a wide range of non-Hermitian research topics and applications, such as adaptive wave control.

In summary, we establish a general paradigm for emergent non-Hermitian topological transition in passive conservative systems through attenuation-gauge duality. By deriving the effective Hamiltonian via the Feshbach projection of the system, we demonstrate that spatially graded coupling to a reservoir generates an imaginary gauge field. This mechanism drives the NHSE and topological phase transitions without the instability risks typically associated with active gain media. This paradigm is broadly applicable and facilitates constructing non-Hermitian platforms in passive and conservative systems and in both linear and nonlinear region. 
The universality of this paradigm allows its implementation across a wide range of physical platforms with high precision, paving the way for scalable and robust non-Hermitian devices in applications such as sensing, energy harvesting, and signal processing (See Supplemental Note Sec.~V,VI).

\textbf{Acknowledgements}
L.X. acknowledges the financial support from NSFC-12574233, GRF-14307422, GRF-14306923, The Chinese University of Hong Kong (CUHK) direct grant 4053582, X. S. acknowledges the financial support from Guangdong Basic and Applied Basic Research Foundation
(Project Nos. 2025B1515020077, 2024A1515030139), NSFC, under the Grant No. T2550093. Correspondence should be addressed to: shenxy66@sysu.edu.cn and xuleixu@cuhk.edu.hk.

\textbf{Competing interests}
The authors declare no competing interest.

\bibliographystyle{apsrev4-2}
\bibliography{reference.bib}

@article{miriLargeAreaSinglemode2012,
	author = {Miri, Mohammad-Ali and LiKamWa, Patrik and Christodoulides, Demetrios N.},
	journal = {Opt. Lett., OL},
	langid = {english},
	month = mar,
	number = {5},
	pages = {764--766},
	publisher = {Optica Publishing Group},
	title = {Large Area Single-Mode Parity--Time-Symmetric Laser Amplifiers},
	volume = {37},
	year = {2012},
}

@article{hodaeiParitytimeSymmetricMicroring2014,
	author = {Hodaei, Hossein and Miri, Mohammad-Ali and Heinrich, Matthias and Christodoulides, Demetrios N. and Khajavikhan, Mercedeh},
	journal = {Science},
	langid = {english},
	month = nov,
	number = {6212},
	pages = {975--978},
	title = {Parity-Time--Symmetric Microring Lasers},
	volume = {346},
	year = {2014},
}

@article{fuModeControlQuasiPT2020,
	author = {Fu, Ting and Wang, Yu-Fei and Wang, Xue-You and Zhou, Xu-Yan and Zheng, Wan-Hua},
	journal = {Chinese Phys. Lett.},
	langid = {english},
	month = apr,
	number = {4},
	pages = {044207},
	publisher = {{Chinese Physical Society and IOP Publishing Ltd}},
	title = {Mode {{Control}} of {{Quasi-PT Symmetry}} in {{Laterally Multi-Mode Double Ridge Semiconductor Laser}}*},
	volume = {37},
	year = {2020},
}

@article{liTopologicalSwitchNonHermitian2020,
	author = {Li, Linhu and Lee, Ching Hua and Gong, Jiangbin},
	journal = {Phys. Rev. Lett.},
	month = jun,
	number = {25},
	pages = {250402},
	publisher = {American Physical Society},
	title = {Topological {{Switch}} for {{Non-Hermitian Skin Effect}} in {{Cold-Atom Systems}} with {{Loss}}},
	volume = {124},
	year = {2020},
}

@article{zhaoNonHermitianTopologicalLight2019,
	author = {Zhao, Han and Qiao, Xingdu and Wu, Tianwei and Midya, Bikashkali and Longhi, Stefano and Feng, Liang},
	journal = {Science},
	langid = {english},
	month = sep,
	number = {6458},
	pages = {1163--1166},
	title = {Non-{{Hermitian}} Topological Light Steering},
	volume = {365},
	year = {2019},
}

@article{shenTopologicalBandTheory2018a,
	author = {Shen, Huitao and Zhen, Bo and Fu, Liang},
	journal = {Phys. Rev. Lett.},
	month = apr,
	number = {14},
	pages = {146402},
	publisher = {American Physical Society},
	title = {Topological {{Band Theory}} for {{Non-Hermitian Hamiltonians}}},
	volume = {120},
	year = {2018},
}

@article{scheibnerNonHermitianBandTopology2020,
	author = {Scheibner, Colin and Irvine, William T. M. and Vitelli, Vincenzo},
	journal = {Phys. Rev. Lett.},
	month = sep,
	number = {11},
	pages = {118001},
	publisher = {American Physical Society},
	title = {Non-{{Hermitian Band Topology}} and {{Skin Modes}} in {{Active Elastic Media}}},
	volume = {125},
	year = {2020},
}

@article{ghatakObservationNonHermitianTopology2020,
	author = {Ghatak, Ananya and Brandenbourger, Martin and {van Wezel}, Jasper and Coulais, Corentin},
	journal = {Proceedings of the National Academy of Sciences},
	month = nov,
	number = {47},
	pages = {29561--29568},
	publisher = {Proceedings of the National Academy of Sciences},
	title = {Observation of Non-{{Hermitian}} Topology and Its Bulk--Edge Correspondence in an Active Mechanical Metamaterial},
	urldate = {2022-09-23},
	volume = {117},
	year = {2020},
}

@article{wangNonHermitianMorphingTopological2022,
	author = {Wang, Wei and Wang, Xulong and Ma, Guancong},
	journal = {Nature},
	langid = {english},
	month = aug,
	number = {7921},
	pages = {50--55},
	title = {Non-{{Hermitian}} Morphing of Topological Modes},
	volume = {608},
	year = {2022},
}

@article{borgniaNonHermitianBoundaryModes2020,
	author = {Borgnia, Dan S. and Kruchkov, Alex Jura and Slager, Robert-Jan},
	issn = {0031-9007, 1079-7114},
	journal = {Phys. Rev. Lett.},
	month = feb,
	number = {5},
	pages = {056802},
	primaryclass = {cond-mat, physics:quant-ph},
	title = {Non-{{Hermitian Boundary Modes}}},
	volume = {124},
	year = {2020},
}

@article{wangTopologicalComplexenergyBraiding2021,
	author = {Wang, Kai and Dutt, Avik and Wojcik, Charles C. and Fan, Shanhui},
	journal = {Nature},
	langid = {english},
	month = oct,
	number = {7879},
	pages = {59--64},
	publisher = {Nature Publishing Group},
	title = {Topological Complex-Energy Braiding of Non-{{Hermitian}} Bands},
	volume = {598},
	year = {2021},
}

@article{scheibnerOddElasticity2020b,
	author = {Scheibner, Colin and Souslov, Anton and Banerjee, Debarghya and Sur{\'o}wka, Piotr and Irvine, William T. M. and Vitelli, Vincenzo},
	journal = {Nat. Phys.},
	langid = {english},
	month = apr,
	number = {4},
	pages = {475--480},
	title = {Odd Elasticity},
	volume = {16},
	year = {2020},
}

@article{veenstraNonreciprocalTopologicalSolitons2024,
	author = {Veenstra, Jonas and Gamayun, Oleksandr and Guo, Xiaofei and Sarvi, Anahita and Meinersen, Chris Ventura and Coulais, Corentin},
	journal = {Nature},
	langid = {english},
	month = mar,
	number = {8004},
	pages = {528--533},
	publisher = {Nature Publishing Group},
	title = {Non-Reciprocal Topological Solitons in Active Metamaterials},
	volume = {627},
	year = {2024},
}

@article{jiangTopologicalInvariantsPhase2018,
	author = {Jiang, Hui and Yang, Chao and Chen, Shu},
	journal = {Phys. Rev. A},
	month = nov,
	number = {5},
	pages = {052116},
	publisher = {American Physical Society},
	title = {Topological Invariants and Phase Diagrams for One-Dimensional Two-Band Non-{{Hermitian}} Systems without Chiral Symmetry},
	volume = {98},
	year = {2018},
}

@article{wangNonHermitianTopologyStatic2023,
	author = {Wang, Aoxi and Meng, Zhiqiang and Chen, Chang Qing},
	journal = {Science Advances},
	month = jul,
	number = {27},
	pages = {eadf7299},
	publisher = {American Association for the Advancement of Science},
	title = {Non-{{Hermitian}} Topology in Static Mechanical Metamaterials},
	volume = {9},
	year = {2023},
}

@article{yangNonHermitianHopflinkExceptional2019,
	author = {Yang, Zhesen and Hu, Jiangping},
	journal = {Phys. Rev. B},
	month = feb,
	number = {8},
	pages = {081102},
	publisher = {American Physical Society},
	title = {Non-{{Hermitian Hopf-link}} Exceptional Line Semimetals},
	volume = {99},
	year = {2019},
}

@article{zhangUniversalNonHermitianSkin2022,
	author = {Zhang, Kai and Yang, Zhesen and Fang, Chen},
	journal = {Nat Commun},
	langid = {english},
	month = may,
	number = {1},
	pages = {2496},
	publisher = {Nature Publishing Group},
	title = {Universal Non-{{Hermitian}} Skin Effect in Two and Higher Dimensions},
	volume = {13},
	year = {2022},
}

@article{guoTheoreticalPredictionNonHermitian2022,
	author = {Guo, Sibo and Dong, Chenxiao and Zhang, Fuchun and Hu, Jiangping and Yang, Zhesen},
	journal = {Phys. Rev. A},
	langid = {english},
	month = dec,
	number = {6},
	pages = {L061302},
	title = {Theoretical Prediction of a Non-{{Hermitian}} Skin Effect in Ultracold-Atom Systems},
	volume = {106},
	year = {2022},
}

@article{zhangEdgeTheoryNonHermitian2024,
	author = {Zhang, Kai and Yang, Zhesen and Sun, Kai},
	journal = {Phys. Rev. B},
	month = apr,
	number = {16},
	pages = {165127},
	publisher = {American Physical Society},
	title = {Edge Theory of Non-{{Hermitian}} Skin Modes in Higher Dimensions},
	volume = {109},
	year = {2024},
}

@misc{liangProbingTopologicalPhase2023a,
	author = {Liang, Jingcheng and Fang, Chen and Hu, Jiangping},
	month = dec,
	number = {arXiv:2401.00530},
	primaryclass = {cond-mat, physics:quant-ph},
	publisher = {arXiv},
	title = {Probing Topological Phase Transition with Non-{{Hermitian}} Perturbations},
	year = {2023},
}

@article{gaoNonHermitianRouteHigherorder2021,
	author = {Gao, He and Xue, Haoran and Gu, Zhongming and Liu, Tuo and Zhu, Jie and Zhang, Baile},
	journal = {Nat Commun},
	langid = {english},
	month = mar,
	number = {1},
	pages = {1888},
	publisher = {Nature Publishing Group},
	title = {Non-{{Hermitian}} Route to Higher-Order Topology in an Acoustic Crystal},
	volume = {12},
	year = {2021},
}

@article{zhouNonHermitianTopologicalMetamaterials2020a,
	author = {Zhou, Di and Zhang, Junyi},
	journal = {Phys. Rev. Res.},
	month = may,
	number = {2},
	pages = {023173},
	publisher = {American Physical Society},
	title = {Non-{{Hermitian}} Topological Metamaterials with Odd Elasticity},
	volume = {2},
	year = {2020},
}

@article{mengNonHermitianTopologicalCoupler2021,
	author = {Meng, Yan and Wu, Xiaoxiao and Shen, Yaxi and Liu, Dong and Liang, Zixian and Zhang, Xiang and Li, Jensen},
	journal = {Sci. China Phys. Mech. Astron.},
	langid = {english},
	month = dec,
	number = {2},
	pages = {224611},
	title = {Non-{{Hermitian}} Topological Coupler for Elastic Waves},
	volume = {65},
	year = {2021},
}

@article{gaoControllingAcousticNonHermitian2024,
	author = {Gao, He and Zhu, Weiwei and Xue, Haoran and Ma, Guancong and Su, Zhongqing},
	journal = {Applied Physics Reviews},
	month = jul,
	number = {3},
	pages = {031410},
	title = {Controlling Acoustic Non-{{Hermitian}} Skin Effect via Synthetic Magnetic Fields},
	volume = {11},
	year = {2024},
}

@article{longNonHermitianTopologicalSystems2022,
	journal = {Phys. Rev. B},
	month = mar,
	number = {10},
	pages = {L100102},
	publisher = {American Physical Society},
	title = {Non-{{Hermitian}} Topological Systems with Eigenvalues That Are Always Real},
	volume = {105},
	year = {2022},
}

@article{zhangAcousticNonHermitianSkin2021a,
	author = {Zhang, Li and Yang, Yihao and Ge, Yong and Guan, Yi-Jun and Chen, Qiaolu and Yan, Qinghui and Chen, Fujia and Xi, Rui and Li, Yuanzhen and Jia, Ding},
	journal = {Nature communications},
	number = {1},
	pages = {6297},
	publisher = {Nature Publishing Group UK London},
	title = {Acoustic Non-{{Hermitian}} Skin Effect from Twisted Winding Topology},
	volume = {12},
	year = {2021}}

@article{imuraGeneralizedBulkedgeCorrespondence2019,
	author = {Imura, Ken-Ichiro and Takane, Yositake},
	journal = {Phys. Rev. B},
	month = oct,
	number = {16},
	pages = {165430},
	publisher = {American Physical Society},
	title = {Generalized Bulk-Edge Correspondence for Non-{{Hermitian}} Topological Systems},
	volume = {100},
	year = {2019},
}

@article{zhangBulkboundaryCorrespondenceNonHermitian2020,
	author = {Zhang, Zhicheng and Yang, Zhesen and Hu, Jiangping},
	journal = {Phys. Rev. B},
	month = jul,
	number = {4},
	pages = {045412},
	publisher = {American Physical Society},
	title = {Bulk-Boundary Correspondence in Non-{{Hermitian Hopf-link}} Exceptional Line Semimetals},
	volume = {102},
	year = {2020},
}

@article{yaoNonHermitianChernBands2018,
	author = {Yao, Shunyu and Song, Fei and Wang, Zhong},
	journal = {Phys. Rev. Lett.},
	month = sep,
	number = {13},
	pages = {136802},
	publisher = {American Physical Society},
	title = {Non-{{Hermitian Chern Bands}}},
	volume = {121},
	year = {2018},
}

@article{yokomizoNonBlochBandTheory2019,
	author = {Yokomizo, Kazuki and Murakami, Shuichi},
	journal = {Phys. Rev. Lett.},
	month = aug,
	number = {6},
	pages = {066404},
	publisher = {American Physical Society},
	title = {Non-{{Bloch Band Theory}} of {{Non-Hermitian Systems}}},
	volume = {123},
	year = {2019},
}

@article{fanReconfigurableTopologicalModes2023,
	author = {Fan, Haiyan and Gao, He and Liu, Tuo and An, Shuowei and Kong, Xianghong and Xu, Guoqiang and Zhu, Jie and Qiu, Cheng-Wei and Su, Zhongqing},
	journal = {Phys. Rev. B},
	month = may,
	number = {20},
	pages = {L201108},
	publisher = {American Physical Society},
	title = {Reconfigurable Topological Modes in Acoustic Non-{{Hermitian}} Crystals},
	volume = {107},
	year = {2023},
}

@article{okumaTopologicalOriginNonHermitian2020a,
	author = {Okuma, Nobuyuki and Kawabata, Kohei and Shiozaki, Ken and Sato, Masatoshi},
	journal = {Phys. Rev. Lett.},
	month = feb,
	number = {8},
	pages = {086801},
	publisher = {American Physical Society},
	title = {Topological {{Origin}} of {{Non-Hermitian Skin Effects}}},
	volume = {124},
	year = {2020},
}

@article{kawabataHigherorderNonHermitianSkin2020,
	author = {Kawabata, Kohei and Sato, Masatoshi and Shiozaki, Ken},
	journal = {Phys. Rev. B},
	month = nov,
	number = {20},
	pages = {205118},
	publisher = {American Physical Society},
	title = {Higher-Order Non-{{Hermitian}} Skin Effect},
	volume = {102},
	year = {2020},
}

@article{zhou2016research,
  title={Research and applications of viscoelastic vibration damping materials: A review},
  author={Zhou, XQ and Yu, DY and Shao, XY and Zhang, SQ and Wang, S},
  journal={Composite Structures},
  volume={136},
  pages={460--480},
  year={2016},
  publisher={Elsevier}
}

@article{dykstra2023buckling,
  title={Buckling metamaterials for extreme vibration damping},
  author={Dykstra, David MJ and Lenting, Coen and Masurier, Alexandre and Coulais, Corentin},
  journal={Advanced materials},
  volume={35},
  number={35},
  pages={2301747},
  year={2023},
  publisher={Wiley Online Library}
}

@article{wang2016tunable,
  title={Tunable digital metamaterial for broadband vibration isolation at low frequency},
  author={Wang, Ziwei and Zhang, Quan and Zhang, Kai and Hu, Gengkai},
  journal={Advanced materials},
  volume={28},
  number={44},
  pages={9857--9861},
  year={2016}
}

@article{ryu2019hybrid,
  title={Hybrid energy harvesters: toward sustainable energy harvesting},
  author={Ryu, Hanjun and Yoon, Hong-Joon and Kim, Sang-Woo},
  journal={Advanced Materials},
  volume={31},
  number={34},
  pages={1802898},
  year={2019},
  publisher={Wiley Online Library}
}

@article{liao2021acoustic,
  title={Acoustic metamaterials: A review of theories, structures, fabrication approaches, and applications},
  author={Liao, Guangxin and Luan, Congcong and Wang, Zhenwei and Liu, Jiapeng and Yao, Xinhua and Fu, Jianzhong},
  journal={Advanced Materials Technologies},
  volume={6},
  number={5},
  pages={2000787},
  year={2021},
  publisher={Wiley Online Library}
}

@article{cao2023observation,
  title={Observation of phononic skyrmions based on hybrid spin of elastic waves},
  author={Cao, Liyun and Wan, Sheng and Zeng, Yi and Zhu, Yifan and Assouar, Badreddine},
  journal={Science Advances},
  volume={9},
  number={7},
  pages={eadf3652},
  year={2023},
  publisher={American Association for the Advancement of Science}
}

@article{chen2021topological,
  title={Topological phononic materials: Computation and data},
  author={Chen, Xing-Qiu and Liu, Jiaxi and Li, Jiangxu},
  journal={The Innovation},
  volume={2},
  number={3},
  year={2021},
  publisher={Elsevier}
}

@article{soneTransitionTopologicalChaotic2025,
  title = {Transition from the Topological to the Chaotic in the Nonlinear {{Su}}--{{Schrieffer}}--{{Heeger}} Model},
  author = {Sone, Kazuki and Ezawa, Motohiko and Gong, Zongping and Sawada, Taro and Yoshioka, Nobuyuki and Sagawa, Takahiro},
  year = 2025,
  month = jan,
  journal = {Nature Communications},
  volume = {16},
  number = {1},
  pages = {422},
}

@article{kawabataHopfBifurcationNonlinear2025,
  title = {Hopf {{Bifurcation}} of {{Nonlinear Non-Hermitian Skin Effect}}},
  author = {Kawabata, Kohei and Nakamura, Daichi},
  year = 2025,
  month = sep,
  journal = {Physical Review Letters},
  volume = {135},
  number = {12},
  pages = {126610},
}

@article{xiaNonlinearTuningPT2021a,
  title = {Nonlinear Tuning of {{PT}} Symmetry and Non-{{Hermitian}} Topological States},
  author = {Xia, Shiqi and Kaltsas, Dimitrios and Song, Daohong and Komis, Ioannis and Xu, Jingjun and Szameit, Alexander and Buljan, Hrvoje and Makris, Konstantinos G. and Chen, Zhigang},
  year = 2021,
  month = apr,
  journal = {Science},
  volume = {372},
  number = {6537},
  pages = {72--76},
  publisher = {American Association for the Advancement of Science},
}

@article{daiNonHermitianTopologicalPhase2023,
  title = {Non-{{Hermitian}} Topological Phase Transitions Controlled by Nonlinearity},
  author = {Dai, Tianxiang and Ao, Yutian and Mao, Jun and Yang, Yan and Zheng, Yun and Zhai, Chonghao and Li, Yandong and Yuan, Jingze and Tang, Bo and Li, Zhihua and Luo, Jun and Wang, Wenwu and Hu, Xiaoyong and Gong, Qihuang and Wang, Jianwei},
  year = 2023,
  month = oct,
  journal = {Nature Physics},
  pages = {1--8},
  publisher = {Nature Publishing Group}
}

@article{okuma2020topological,
  title={Topological origin of non-Hermitian skin effects},
  author={Okuma, Nobuyuki and Kawabata, Kohei and Shiozaki, Ken and Sato, Masatoshi},
  journal={Physical review letters},
  volume={124},
  number={8},
  pages={086801},
  year={2020},
  publisher={APS}
}

@article{zhang2020correspondence,
  title={Correspondence between winding numbers and skin modes in non-Hermitian systems},
  author={Zhang, Kai and Yang, Zhesen and Fang, Chen},
  journal={Physical Review Letters},
  volume={125},
  number={12},
  pages={126402},
  year={2020},
  publisher={APS}
}

@article{nie2025designing,
  title={Designing unique mechanical modes through an extension of the quantum hopping method},
  author={Nie, Haoran and Shen, Xiangying and Xu, Lei},
  journal={Proceedings of the National Academy of Sciences},
  volume={122},
  number={46},
  pages={e2423603122},
  year={2025},
  publisher={National Academy of Sciences}
}

\end{document}


\title{Supplemental Material}
\author{Haoran Nie}
\author{Chaoran Jiang}
\author{Xiangying Shen}
\author{Lei Xu}
\maketitle

\section{I. Theoretical Derivation of the Emergent Non-Hermitian Skin Effect via Gradient-Gauge Duality}

This section provides a general derivation of how effective non-Hermitian dynamics can arise in a passive conservative system through coupling to a structured reservoir. We establish the Gradient-Gauge Duality framework, demonstrating that a spatial gradient in the system-reservoir coupling is mathematically dual to an emergent imaginary gauge field in the momentum space, driving the non-Hermitian skin effect (NHSE).

\subsection{A. Effective Hamiltonian via Feshbach Projection}

We consider a closed conservative total system $W=S\oplus R$ composed of a primary subsystem $S$ and a reservoir $R$. The total Hamiltonian $H_{\mathrm{tot}}$ is Hermitian and can be written in block form as
\begin{equation}
H_{\mathrm{tot}}=
\begin{pmatrix}
H_{SS} & H_{SR}\\
H_{RS} & H_{RR}
\end{pmatrix},
\end{equation}
where $H_{SS}$ and $H_{RR}$ govern the internal dynamics of the subsystem and reservoir, and $H_{SR}=H_{RS}^\dagger$ describes their interaction. The total eigenvalue equation at frequency $\omega$ is $(H_{\mathrm{tot}}-\omega I)\Psi=0$, with $\Psi=(\psi_S,\psi_R)^T$.

To obtain an effective description for $S$, we expand the reservoir component,
\begin{equation}
H_{RS}\psi_S+H_{RR}\psi_R=\omega \psi_R,
\end{equation}
and solve
\begin{equation}
\psi_R=G_R(\omega)H_{RS}\psi_S, \qquad
G_R(\omega)=(\omega I-H_{RR})^{-1}.
\end{equation}
Substituting back yields
\begin{equation}
\big[H_{SS}+\Sigma(\omega)\big]\psi_S=\omega \psi_S,
\end{equation}
with the self-energy operator
\begin{equation}
\Sigma(\omega)=H_{SR}G_R(\omega)H_{RS}.
\end{equation}

For an extended reservoir with a continuous spectrum, the physically relevant quantity is the retarded Green's function
\begin{equation}
G_R^{r}(\omega)=\lim_{\epsilon\to 0^+}\big[(\omega+i\epsilon)I-H_{RR}\big]^{-1},
\end{equation}
which defines the retarded self-energy $\Sigma^{r}(\omega)=H_{SR}G_R^{r}(\omega)H_{RS}$ and the corresponding effective Hamiltonian $H_{\mathrm{eff}}^{r}(\omega)=H_{SS}+\Sigma^{r}(\omega)$. Passivity implies that the anti-Hermitian part of $\Sigma^{r}$ is negative semi-definite,
\begin{equation}
\frac{\Sigma^{r}(\omega)-\Sigma^{r\dagger}(\omega)}{2i}\le 0,
\end{equation}
or equivalently $\mathrm{Im}\,\Sigma^{r}(\omega)\le 0$ in the scalar case. This condition encodes energy outflow from $S$ into $R$ in the absence of internal sources.

\subsection{B. Microscopic Derivation of a Gradient Self-Energy}

We next derive the form of $\Sigma^{r}(\omega)$ under a structured-reservoir assumption. The matrix element between two subsystem sites $i$ and $j$ reads
\begin{equation}
\Sigma^{r}_{ij}(\omega)=\sum_{m,n}(H_{SR})_{im}\,(G_R^{r})_{mn}(\omega)\,(H_{RS})_{nj},
\end{equation}
where $m,n$ label reservoir degrees of freedom.

We assume that each site $i$ couples primarily to a local set of reservoir degrees of freedom $r_i$, so that $(H_{SR})_{im}=v_i\,\delta_{m,r_i}$ with a site-dependent coupling strength $v_i$. We further assume a short reservoir correlation length compared with the gradient scale, so that in the local basis
\begin{equation}
(G_R^{r})_{r_i,r_j}(\omega)\approx g_R^{r}(\omega)\,\delta_{r_i,r_j},
\end{equation}
which reduces to $g_R^{r}(\omega)\delta_{ij}$ when $r_i$ is uniquely associated with site $i$. These approximations yield
\begin{equation}
\Sigma^{r}_{ij}(\omega)\approx v_i v_j^* g_R^{r}(\omega)\,\delta_{ij},
\end{equation}
and thus a predominantly local self-energy
\begin{equation}
\Sigma^{r}_{ii}(\omega)\approx |v_i|^2 g_R^{r}(\omega).
\end{equation}

Writing the local response as
\begin{equation}
g_R^{r}(\omega)=\Lambda(\omega)-i\mathcal{D}(\omega),
\qquad \mathcal{D}(\omega)\ge 0,
\end{equation}
we obtain
\begin{equation}
\Sigma^{r}_{ii}(\omega)=\Delta_i(\omega)-i\Gamma_i(\omega),
\qquad
\mathbf{\Gamma_i(\omega)=|v_i|^2\mathcal{D}(\omega).}
\end{equation}
Crucially, the effective decay rate depends quadratically on the coupling strength, $\Gamma_i\propto |v_i|^2$. Thus, a designed exponential gradient in the system-reservoir coupling,
\begin{equation}
v(x)=v_0 e^{\eta x},
\end{equation}
generates a decay profile with a doubled gradient exponent:
\begin{equation}
\Gamma(x)=\Gamma_0 e^{2\eta x,
\qquad
\Gamma_0=|v_0|^2\mathcal{D}(\omega).}
\end{equation}
In the following, we show that such a gradient can bias stationary spatial envelopes and produce pronounced non-Bloch accumulation in finite systems.

\subsection{C. WKB Analysis and Local Gradient-Gauge Duality}

We employ a WKB analysis to establish the local duality between the gradient decay and an imaginary momentum. We consider the continuum limit and an adiabatic gradient ($\eta a\ll 1$ for lattice spacing $a$). We also assume weak non-Hermiticity such that the imaginary part of the local wave vector satisfies $|\kappa a|\ll 1$.

We take the ansatz
\begin{equation}
\psi(x)\sim \exp\!\left(i\int^x k(x')\,dx'\right),
\qquad
k(x)=k_R(x)+i\kappa(x),
\end{equation}
where $k_R(x)$ is the local real momentum and $\kappa(x)$ sets the inverse envelope length. Substituting into the effective equation and expanding to leading order in $\kappa$, the local dispersion can be written as
\begin{equation}
\omega \approx \epsilon(k_R)
+ i\,v_g(k_R)\,\kappa(x)
- i\,\Gamma(x),
\end{equation}
where $\epsilon(k)$ is the dispersion of the isolated subsystem and $v_g=\partial\epsilon/\partial k$ is the group velocity.

Although $H_{\mathrm{eff}}^{r}$ is non-Hermitian, the stationary patterns correspond to subsystem projections of real-frequency excitations of the full Hermitian system $W$. Taking the imaginary part at real $\omega$ yields
\begin{equation}
0 \approx v_g(k_R)\,\kappa(x)-\Gamma(x),
\end{equation}
so that
\begin{equation}
\kappa(x)\approx \frac{\Gamma(x)}{v_g(k_R)}.
\label{eq:si_kappa_gamma}
\end{equation}
This equation represents the local Gradient-Gauge Duality: the spatial decay gradient $\Gamma(x)$ is dynamically equivalent to a local imaginary momentum shift $\kappa(x)$.

Accordingly, the mode envelope is given by:
\begin{equation}
|\psi(x)|\sim
\exp\!\left(-\int^x \kappa(x')\,dx'\right)
\approx
\exp\!\left[-\int^x \frac{\Gamma(x')}{v_g(k_R)}\,dx'\right].
\label{eq:si_wkb_envelope_general}
\end{equation}

Substituting the explicit gradient profile $\Gamma(x) = \Gamma_0 e^{2\eta x}$, yields the closed-form WKB envelope:
\begin{equation}
|\psi(x)| \propto
\exp\!\left[-\frac{\Gamma_0}{2\eta\,v_g(k_R)}\left(e^{2\eta x}-1\right)\right].
\label{eq:si_wkb_envelope_closed}
\end{equation}
Consequently, the boundary envelope ratio scales as
\begin{equation}
\frac{|\psi(L)|}{|\psi(0)|}
\sim
\exp\!\left[-\frac{\Gamma_0}{2\eta\,v_g(k_R)}\left(e^{2\eta L}-1\right)\right].
\label{eq:si_boundary_ratio}
\end{equation}

\subsection{D. Emergent Gauge Field and Topological Classification}

To rigorously characterize the topological phase, we apply a coarse-graining procedure to the local duality. We define an effective imaginary gauge field $\bar{\lambda}$ such that a hypothetical uniform system with this parameter would produce the same net accumulation $\Xi(L) \equiv e^{-\bar{\lambda}L}$. This yields:
\begin{equation}
\bar{\lambda} = \frac{1}{L}\int_0^L \kappa(x)\,dx \approx \frac{1}{L v_g} \int_0^L \Gamma(x)\,dx.
\end{equation}
This emergent gauge field $\bar{\lambda}$ captures the global tendency of the skin effect. We can now construct a translationally invariant reference Hamiltonian, $h_{\text{ref}}(k)$, by applying the imaginary gauge transformation to the Hermitian dispersion $\epsilon(k)$:
\begin{equation}
h_{\text{ref}}(k) \equiv \epsilon(k - i\bar{\lambda}).
\end{equation}
For a nearest-neighbor tight-binding chain with dispersion $2t\cos k$, this yields:
\begin{equation}
h_{\text{ref}}(k) = t(e^{\bar{\lambda}}e^{ik} + e^{-\bar{\lambda}}e^{-ik}).
\end{equation}
The associated point-gap winding number with respect to a gap center energy $E_{\mathrm{C}}$ is
\begin{equation}
w=\frac{1}{2\pi i}\int_{-\pi}^{\pi}\frac{d}{dk}
\ln\det\big[H_{\mathrm{ref}}(k)-E_{\mathrm{C}}\big]\,dk.
\end{equation}
This confirms that the passive gradient system falls into the same topological class as the non-reciprocal Hatano-Nelson model, governed by the emergent gauge field $\bar{\lambda}$.

\subsection{E. Mechanism of Skin-Mode Reversal in Coupled Chains}

We now address the coupled-chain implementation discussed in the main text. The system consists of two chains, $\alpha$ and $\beta$, coupled by $t_\perp$. The structured reservoir induces distinct effective attenuation profiles $\Gamma_\alpha(x)$ and $\Gamma_\beta(x)$, which map to competing effective gauge fields $\bar{\lambda}_\alpha$ and $\bar{\lambda}_\beta$.

The full reference Hamiltonian for the coupled system is constructed by assembling the effective single-chain components into a $2\times2$ matrix in the basis $(\psi_{\alpha},\psi_{\beta})^{T}$:
\begin{equation}
H_{ref}^{2\times 2}(k) = \begin{pmatrix}
h_{\alpha}^{ref}(k) & t_\perp \\
t_\perp & h_{\beta}^{ref}(k)
\end{pmatrix}
= \begin{pmatrix}
t(e^{\bar{\lambda}_\alpha}e^{ik} + e^{-\bar{\lambda}_\alpha}e^{-ik}) & t_\perp \\
t_\perp & t(e^{\bar{\lambda}_\beta}e^{ik} + e^{-\bar{\lambda}_\beta}e^{-ik})
\end{pmatrix}.
\label{eq:Href_2x2}
\end{equation}
To understand the topological phase transition, we analyze the global spectral topology. The condition for point-gap closing at $E_{C}=0$ is:
\begin{equation}
\det\!\left[H_{\text{ref}}^{2\times 2}(k)\right] = h_{\alpha}^{\text{ref}}(k)\,h_{\beta}^{\text{ref}}(k)-t_\perp^2 = 0,
\end{equation}
for some momentum $k$.

In the weak-coupling regime ($t_\perp \ll t_c$), the spectrum consists of two loops, and the global topology is non-trivial ($w=1$) due to the asymmetry $|\bar{\lambda}_\alpha| \neq |\bar{\lambda}_\beta|$. As $t_\perp$ increases across a critical threshold $t_c$, the point gap closes and reopens with an inverted spectral topology ($w=-1$). The winding number is defined as:
\begin{equation}
w = \frac{1}{2\pi i} \oint_{-\pi}^{\pi} \frac{d}{dk} \ln \det [H_{ref}^{2\times 2}(k)] dk.
\label{eq:winding}
\end{equation}
This confirms that the skin mode reversal is a topological phase transition driven by the competition between the inter-chain coupling and the emergent gauge fields.
  
\section{F. Microscopic Justification: Adiabatic Continuity from Open Outflow to Conservative Localization}

\begin{figure}[h]
    \centerline{\includegraphics[width=0.95\linewidth]{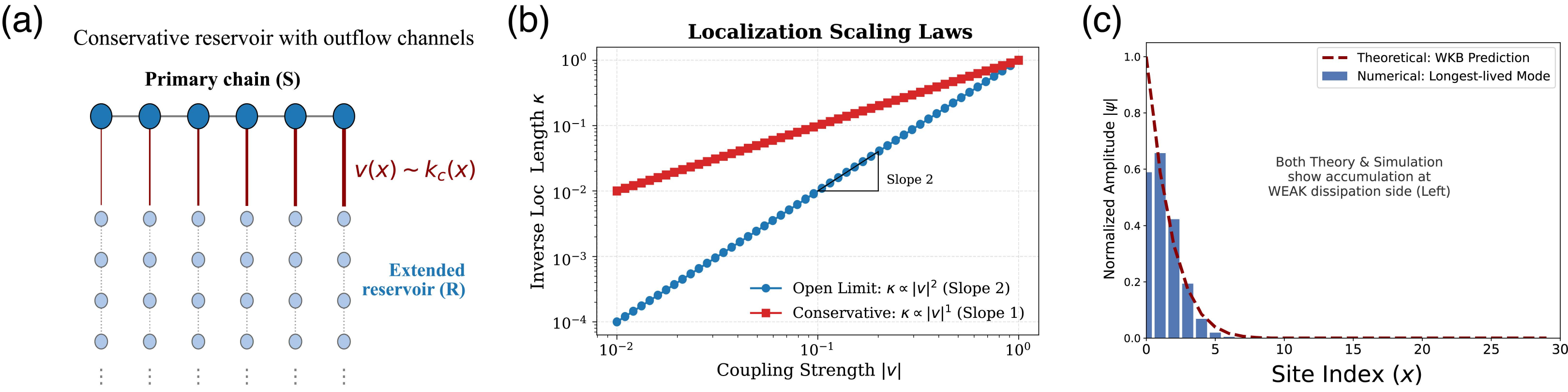}}
    \begin{flushleft}\label{fig:S1}FIG.S1. Lossless outflow reservoir model based on a semi-infinite spring chain.
  (a) Schematic: the subsystem couples locally to an extended, conservative reservoir that supports outflow channels.
  (b) At a fixed working frequency $\omega_0$, the induced outflow rate satisfies
  $\Gamma(x,\omega_0)=|v(x)|^2\mathcal{D}(\omega_0)$. Note that due to this quadratic dependence, an exponential coupling gradient $v(x)\propto e^{\eta x}$ yields a decay profile $\Gamma(x)\propto e^{2\eta x}$.
  (c) Comparison of the WKB prediction and Numerical calculation of the single chain skin mode. \end{flushleft}
  \end{figure}
  
This section provides a rigorous microscopic derivation bridging the ideal non-Hermitian skin effect (arising in open systems) and the conservative localization observed in our experimental setup (finite systems). By analyzing the self-energy corrections in two distinct limits---purely dissipative and purely reactive---we demonstrate that while the scaling law of the localization length changes, the direction of skin accumulation remains topologically robust, protected by the gradient geometry. This establishes the adiabatic continuity between the two regimes.

\noindent\textbf{The Open Dissipative Limit -- Outflow-Driven Decay}
We first consider the ideal case where the subsystem couples to an extended, reflectionless reservoir, modeled as a semi-infinite mass-spring chain with masses $m_0$ and springs $k_0$ as illustrated in Fig.S1(a). The surface retarded Green's function (compliance) $g_{\mathrm{surf}}^r(\omega)$ accounts for the boundary response of the reservoir. Utilizing the self-similarity of the semi-infinite chain, $g_{\mathrm{surf}}^r$ satisfies the continued-fraction equation:
\begin{equation}
g_{\mathrm{surf}}^r(\omega) = \frac{1}{k_0 - m_0(\omega+i0^+)^2 - k_0^2 g_{\mathrm{surf}}^r(\omega)}.
\end{equation}
Solving this quadratic equation yields the physically retarded branch. Crucially, within the reservoir's passband ($0 < \omega < 2\sqrt{k_0/m_0}$), $g_{\mathrm{surf}}^r$ acquires a non-zero imaginary part:
\begin{equation}
g_{\mathrm{surf}}^r(\omega) = \Lambda(\omega) - i \mathcal{D}(\omega), \quad \text{with } \mathcal{D}(\omega) > 0.
\end{equation}
Here, $\mathcal{D}(\omega)$ represents the spectral density of states into which energy escapes irreversibly to infinity. The effective self-energy term added to the subsystem Hamiltonian is:
\begin{equation}
\Sigma(x, \omega) \approx |v(x)|^2 g_{\mathrm{surf}}^r(\omega) = \Delta(x) - i \Gamma(x),
\end{equation}
where the imaginary part $\Gamma(x) = |v(x)|^2 \mathcal{D}(\omega)$ acts as an effective dissipation rate. We apply the WKB ansatz $\psi(x) \sim e^{i \int (k_R + i\kappa) dx}$ to the effective dispersion relation $\omega \approx \epsilon(k_R) + i v_g \kappa - i \Gamma$. Balancing the imaginary terms yields the standard non-Hermitian skin effect result:
\begin{equation}
v_g \kappa_{\mathrm{NH}}(x) \approx \Gamma(x) \implies \kappa_{\mathrm{NH}}(x) \approx \frac{\mathcal{D}(\omega)}{v_g} |v(x)|^2.
\end{equation}
\textit{Result A:} In the open limit, the localization rate $\kappa$ scales quadratically with the coupling strength, $\kappa \propto |v(x)|^2$, driven by the gradient of dissipation.

\noindent\textbf{The Conservative Reactive Limit -- Potential-Driven Squeezing}
In our actual experiment, the reservoir consists of finite grounded springs. The energy outflow is suppressed ($\mathcal{D}(\omega) \to 0$) because the waves are fully reflected by the rigid ground. The self-energy becomes predominantly real:
\begin{equation}
\Sigma(x, \omega) \approx \Delta(x) = |v(x)|^2 \mathrm{Re}[g_{\mathrm{surf}}^r(\omega)].
\end{equation}
Physically, this real self-energy acts as a position-dependent onsite potential $V_{\mathrm{eff}}(x) \equiv \Delta(x)$. The system remains Hermitian, but the translational symmetry is broken by this gradient potential. To find the spatial envelope, we analyze the real part of the local dispersion relation:
\begin{equation}
\omega = \epsilon(k) + V_{\mathrm{eff}}(x).
\end{equation}
Localization occurs when the system is locally pushed into a band gap (classically forbidden region). We expand the band dispersion $\epsilon(k)$ near a band edge (e.g., $k=0$) where the group velocity vanishes, using the effective mass approximation $\epsilon(k) \approx E_0 + \alpha k^2$:
\begin{equation}
\omega \approx E_0 + \alpha (k_R + i\kappa)^2 + \Delta(x).
\end{equation}
In the localization regime, we set $k_R=0$ (purely evanescent solution), leading to:
\begin{equation}
\omega \approx E_0 - \alpha \kappa^2 + \Delta(x).
\end{equation}
Solving for the inverse decay length $\kappa(x)$:
\begin{equation}
\alpha \kappa^2 \approx \Delta(x) - (\omega - E_0) \implies \kappa_{\mathrm{H}}(x) \approx \sqrt{\frac{\Delta(x) - \delta E}{\alpha}}.
\end{equation}
Substituting $\Delta(x) \propto |v(x)|^2$, we obtain:
\begin{equation}
\kappa_{\mathrm{H}}(x) \propto \sqrt{|v(x)|^2} = |v(x)|.
\end{equation}
\textit{Result B:} In the conservative limit, the localization rate scales linearly with the coupling strength, $\kappa \propto |v(x)|$, driven by the gradient of the effective potential (wavefunction squeezing).
In Fig.S1(b), we plot the scaling Laws of the open limit and conservative limit, respectively. The parameter dependencies are clearly exhibited. In Fig.S1(c), we calculate the non-Hermitian skin mode of a 1D chain using WKB prediction and numerical calculation. It can be clearly observed that the two results are in good agreement. 

\subsection{3. Topological Robustness via Adiabatic Connectivity: A Unified Dispersion Model}

To rigorously prove that the experimentally observed localization is topologically equivalent to the intrinsic non-Hermitian skin effect, we must demonstrate a smooth deformation between the two regimes without closing the point gap. Adiabatic connectivity is simply the topological homotopy of the Hamiltonian under a specified path transformation. A subtle challenge arises because the two limits operate in different kinematic regimes: the dissipative limit typically involves mid-band transport ($v_g \neq 0$), while the conservative localization inherently occurs near a band edge ($v_g \to 0$) where the group velocity vanishes.

To bridge these regimes rigorously, we employ a unified dispersion model that retains both first-order (group velocity) and second-order (effective mass) terms in the $k\cdot p$ expansion, explicitly including the energy detuning $\delta E \equiv \omega - E_0$. We parameterize the reservoir's response phase by $\phi \in [0, \pi/2]$, such that the self-energy is $\Sigma(\phi) \propto e^{-i\phi}$. The generalized dispersion relation is:
\begin{equation}
-\alpha \kappa^2 + i v_g(k_R) \kappa + \Sigma(\phi) - \delta E \approx 0
\label{eq:unified_dispersion_full}
\end{equation}
where $v_g$ is the residual group velocity (which vanishes in the strict band-edge limit), $\alpha$ is the band curvature (inverse effective mass), and $\kappa$ is the inverse localization length. The solution for $\kappa$ is given by the quadratic formula:
\begin{equation}
\kappa(\phi) = \frac{iv_g \pm \sqrt{-v_g^2 + 4\alpha (\Sigma(\phi) - \delta E)}}{2\alpha}
\label{eq:kappa_solution}
\end{equation}
This unified solution naturally interpolates between the two limiting scaling laws derived in the previous subsections:

\textbf{Dissipative Limit ($\phi = \pi/2$):}
In the open system limit where the mode is away from the band edge ($|v_g|^2 \gg |4\alpha \Sigma|$), we can expand the square root in Eq. (\ref{eq:kappa_solution}) to first order. The leading term yields:
\begin{equation}
\kappa \approx \frac{\Sigma(\pi/2)}{-i v_g} = \frac{-i |v|^2 \mathcal{D}}{-i v_g} = \frac{\mathcal{D}}{v_g} |v|^2
\end{equation}
This recovers the quadratic scaling $\kappa \propto |v|^2$ characteristic of the standard non-Hermitian skin effect.

\textbf{Conservative Limit ($\phi = 0$):}
In the closed system limit, localization forces the state to the band edge where $v_g \to 0$. The first-order term in Eq. (\ref{eq:unified_dispersion_full}) vanishes, and the solution is dominated by the second-order term:
\begin{equation}
\kappa \approx \sqrt{\frac{\Sigma(0) - \delta E}{\alpha}} = \sqrt{\frac{\Delta - \delta E}{\alpha}} \propto \sqrt{|v|^2} = |v|
\end{equation}
This recovers the linear scaling $\kappa \propto |v|$ characteristic of potential-gradient localization.

\paragraph{Proof of Directional Robustness via Adiabatic Continuity.}
To rigorously demonstrate that the observed localization is topologically protected, we prove that the non-Hermitian point gap remains open throughout the adiabatic deformation. In our effective theory, a point-gap closing corresponds to the decay rate $\kappa$ crossing the imaginary axis (i.e., $\mathrm{Re}(\kappa) \to 0$), which signifies a transition from a localized skin mode to a delocalized propagating state.

We proceed by contradiction. Assume that at some intermediate phase $\phi \in [0, \pi/2]$, the localization vanishes, implying $\kappa$ becomes purely imaginary: $\kappa = i q$ with $q \in \mathbb{R}$. This corresponds to a real Bloch wavevector $k = k_R - q$. Substituting this ansatz into Eq.~(\ref{eq:unified_dispersion_full}):
\begin{equation}
\alpha q^2 - v_g q + \Sigma(\phi) - \delta E = 0.
\end{equation}
Since $\alpha$, $v_g$, $q$, and $\delta E$ are real parameters derived from the band structure and excitation frequency, the imaginary part of this equation must vanish:
\begin{equation}
\mathrm{Im}[\Sigma(\phi)] = 0.
\end{equation}
This condition leads to a contradiction in both limits:
\textbf{Case $\phi > 0$ (Non-Hermitian Regime):} The presence of reservoir outflow ensures $\mathrm{Im}[\Sigma(\phi)] \propto -\sin\phi \neq 0$. Thus, the equation cannot be satisfied, and no propagating solution ($\mathrm{Re}(\kappa)=0$) exists. The mode must remain localized.
    
textbf{Case $\phi = 0$ (Hermitian Limit):} The self-energy becomes real ($\Sigma \to \Delta$). The equation reduces to $\alpha q^2 - v_g q + (\Delta - \delta E) = 0$. For conservative localization to occur, the system must operate within a band gap or classically forbidden region. In the band-edge limit ($v_g \to 0$), this requires the potential step to exceed the energy detuning: $(\Delta - \delta E)/\alpha > 0$. Under this condition, the quadratic equation for $q$ yields no real solutions for $q$, meaning a propagating state is physically forbidden.

Since $\mathrm{Re}(\kappa)$ is strictly non-zero and continuous for all $\phi \in [0, \pi/2]$, the global accumulation factor $\bar{\lambda} \propto \int \kappa dx$ never crosses zero. Consequently, the reference Hamiltonian $H_{\text{ref}}(k)$ maintains an open point gap throughout the deformation. The spectral winding number $w$, being a topological invariant of the open-gap phase, remains constant. This proves that the potential-driven localization in our passive experiment is the adiabatic continuation of the dissipation-driven non-Hermitian skin effect, sharing the same topological class.

\section{II. CONSTRUCTION OF THE MECHANICAL DYNAMICAL MATRIX}

\begin{figure}[h]
    \centerline{\includegraphics[width=0.8\linewidth]{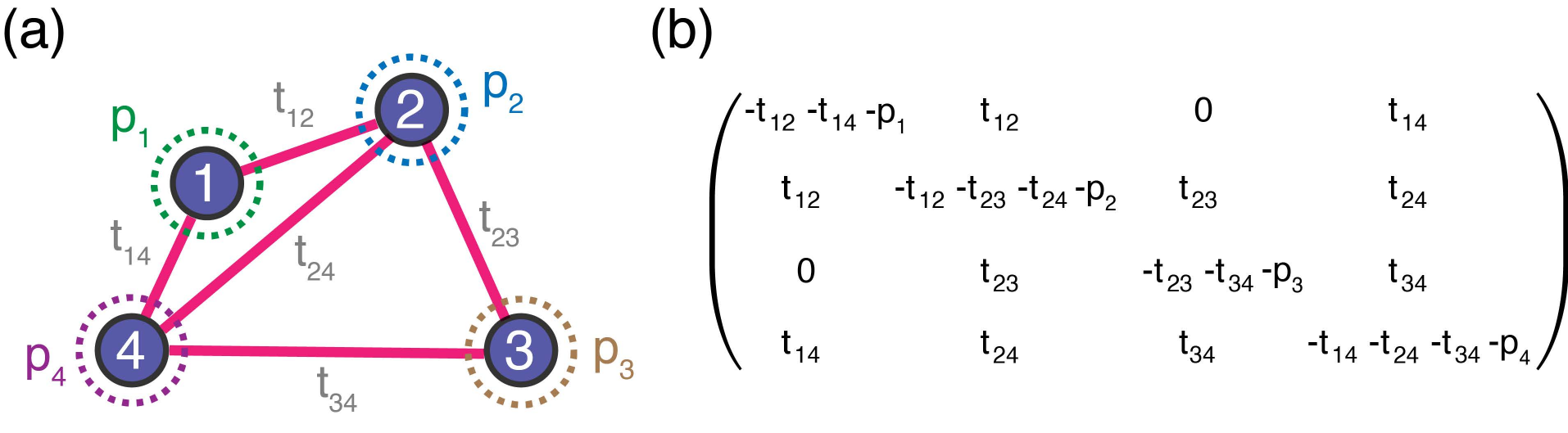}}
    \begin{flushleft}\label{fig:S2}FIG. S2. Schematic of the construction principle for the mechanical dynamical matrix using the hopping method. (a) A generic 2D network with 4 nodes connected by springs. (b) The corresponding dynamical matrix $D$. Note that each spring connection $k_{ij}$ contributes to both the off-diagonal hopping terms (interaction) and the diagonal onsite terms (reaction forces). In our double-chain model, this principle implies that the grounded vertical springs modify the diagonal onsite potentials without altering the inter-site hopping, allowing for the independent engineering of the gradient self-energy $\Sigma(x)$.\end{flushleft}
  \end{figure}

\subsection*{A. Dynamical matrix}
  
In this section, we derive the dynamical matrix for the coupled mass-spring network and establish its mapping to the effective tight-binding model used in the main text. The system consists of $N$ unit cells, each containing two masses $m$ (labeled $\alpha$ and $\beta$). The masses are connected by intra-chain springs with stiffness $k_t$, inter-chain springs with stiffness $k_\perp$, and are grounded to a rigid base via vertical springs with stiffness $k_{g,\alpha}^{(n)}$ and $k_{g,\beta}^{(n)}$ at the $n$-th unit cell.

The equation of motion for the displacement $u_{n,j}$ (where $j \in \{\alpha, \beta\}$) follows Newton's second law. Assuming harmonic motion with frequency $\omega$, the force balance yields the eigenvalue equation for the dynamical matrix $D$:
\begin{equation}
-\omega^2 u_{n,\alpha} = -k_t (2u_{n,\alpha} - u_{n+1,\alpha} - u_{n-1,\alpha}) - k_\perp (u_{n,\alpha} - u_{n,\beta}) - k_{g,\alpha}^{(n)} u_{n,\alpha}.
\end{equation}
A similar equation holds for the $\beta$ chain. By setting the mass $m=1$, the dynamical matrix $D$ takes the form of a tight-binding Hamiltonian. We identify the hopping parameters as $t = k_t$ and $t_\perp = k_\perp$.

A distinguishing feature of mechanical networks is that the diagonal terms (onsite potentials) naturally include contributions from all connected springs due to the reaction forces (Newton's third law). This construction principle, where a spring connection $k_{ij}$ contributes to both the off-diagonal hopping $D_{ij}$ and the diagonal elements $D_{ii}$ and $D_{jj}$, is schematically illustrated for a generic node network in Fig.~S2. Specifically for our double chain, the total diagonal stiffness at node $(n, \alpha)$ is $K_{ii} = 2k_t + k_\perp + k_{g,\alpha}^{(n)}$. To map this to the theoretical form $H = \sum [t c^\dagger_i c_j + V_i c^\dagger_i c_i]$, the onsite potential $V_{n,\alpha}$ corresponds to the total effective stiffness of the site.

Crucially, the grounded vertical springs $k_{g,j}^{(n)}$ allow for the independent tuning of the onsite potential without altering the inter-site hopping terms. In our design, we engineer the stiffness $k_{g,j}^{(n)}$ to vary exponentially along the chain index $n$. This spatial gradient in the grounded stiffness serves as the physical implementation of the gradient coupling $v(x)$ to the reservoir discussed in Section I. Consequently, the onsite diagonal terms of the dynamical matrix acquire a spatial dependence $V_{n,j} \approx V_0 + \delta V_j e^{\eta n}$. Through the Feshbach projection mechanism derived in Sec.~I, a spatial gradient in the grounded stiffness
implements a spatially structured coupling between $S$ and auxiliary degrees of freedom, yielding a
position-dependent self-energy $\Sigma_j^r(x,\omega)=\Delta_j(x,\omega)-i\Gamma_j(x,\omega)$ in the reduced description.
The anti-Hermitian component $\Gamma_j$ is nonzero when the effective reservoir supports outflow channels
(e.g., an extended reservoir with a dense/continuous spectrum at the working frequency; see Sec.~I.F and Fig.~S2),
whereas in the purely reactive finite limit $\Gamma_j\to 0$ and $\Sigma_j^r$ becomes predominantly real.

\subsection*{B. System construction}
The system consists of two parallel chains of coupled oscillators. The masses are connected by horizontal coil springs providing intra-chain hopping $t$ and inter-chain coupling $t_\perp$. The structured reservoir is implemented by grounded vertical springs, whose stiffness is spatially graded to induce the effective non-Hermitian self-energy.

\textbf{1. Tuning Mechanism via Spring Geometry}
The tunable coupling $t_\perp$ is achieved by adjusting the effective length of the connecting springs. For a spring connecting two nodes $i$ and $j$ with an equilibrium length $l$, the restoring force under a small transverse displacement $u_j - u_i$ is given by $F_{j\to i} \approx F_0 (u_j - u_i)/l$, where $F_0$ is the static tension. Consequently, the effective hopping parameter $t_{ij}$ scales inversely with the length:
\begin{equation*}
t_{ij} \approx \frac{F_0}{l}.
\end{equation*}
In our setup, we utilized springs with a wire diameter of \SI{0.3}{mm} and a stiffness of \SI{179.13}{N/m}. By adjusting the movable clamps, the inter-chain coupling $t_\perp$ was continuously tuned over a range of \SI{19}{N/m} to \SI{58}{N/m}. The intra-chain coupling $t$ was fixed at approximately \SI{30}{N/m}. The onsite potential gradients were set to $\nabla_\alpha = 0.04$ and $\nabla_\beta = -0.02$ (normalized units) to establish the competitive outflow (attenuation) landscape required for the skin mode reversal.

\textbf{2. Measurement Protocol}
The system was excited by a mechanical shaker applying a sinusoidal force at one boundary. The excitation frequency was set to \SI{26.57}{Hz}, corresponding to the fundamental mode of the coupled system where the non-Hermitian skin effect is most pronounced due to the maximal spectral overlap with the reservoir continuum. The steady-state out-of-plane displacement of each oscillator was recorded using a CMOS laser displacement sensor (Keyence GC05-30NW) with a spatial resolution of \SI{10}{\micro\meter} and a sampling response time of \SI{1.5}{\milli\second}. The spatial mode profiles shown in Fig.~3 of the main text were reconstructed by sequentially measuring the amplitude at each lattice site for three distinct coupling regimes: $t_\perp$ well below, near, and well above the critical point $t_c$.

\section{III. EXTENSION TO TWO DIMENSIONAL SYSTEM}
\begin{figure*}[htpb]
\renewcommand{\thefigure}{S3}
\includegraphics[width=0.6\linewidth]{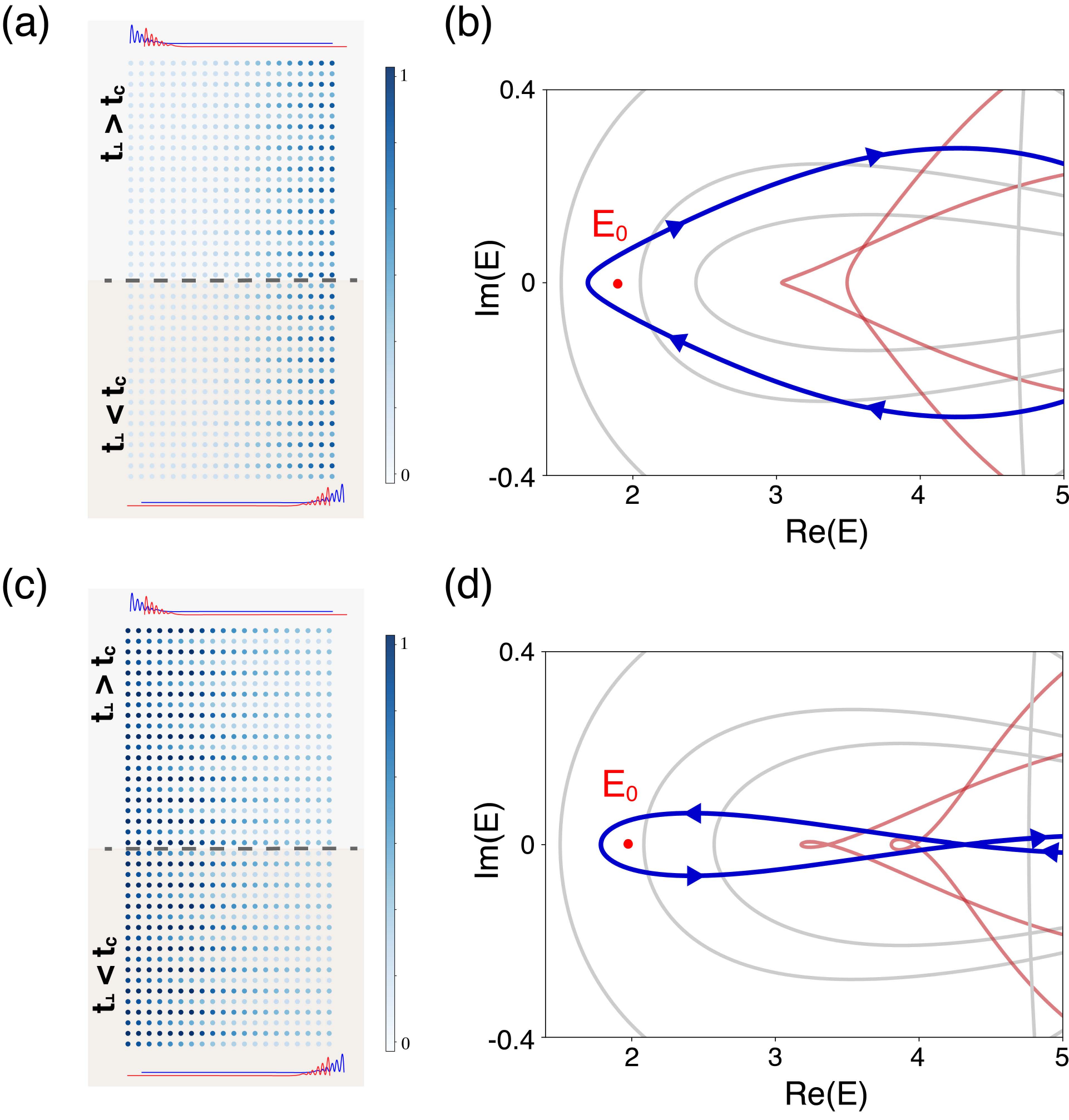}
\caption{\label{fig:S3} Extension to 2D systems by stacking the non-Hermitian chains. (a) Mode distribution when the top zone $t_\perp>t_c$ and bottom zone $t_\perp<t_c$ while bottom zone dominates. (b) exhibits the spectrum winding relation of (a). Grey lines correspond to the floppy modes of the 2D finite boundaries, which is negligible. Blue line is the low energy winding band with winding direction indicated by arrows. (c) is reversed mode distribution when the top zone $t_\perp>t_c$ and bottom zone $t_\perp<t_c$ while top zone dominates. (d) illustrates the spectrum of system in (c). }
\end{figure*}

As demonstrated in the main text, we construct a two-dimensional system by stacking the one-dimensional double chains directly along the $y$-direction. The top and bottom halves belong to distinct regions: $t_\perp > t_c$ in the top and $t_\perp < t_c$ in the bottom, coupling the two regions using either the top or bottom $t_\perp$ (with results remaining insensitive to the choice of this coupling). In isolation, these two regions would support skin modes localized at opposite boundaries, as indicated by the curves on the top and bottom. Surprisingly and counter-intuitively, the skin modes of this 2D system cannot be described as a simple linear superposition of the individual partitions. Instead, they manifest as a collective entity localized at a specific boundary. This anomalous phenomenon can be elucidated using an approach analogous to that employed for the 1D chain. We treat the 2D system as a supercell along the $y$ direction to compute the corresponding spectrum. By examining the winding properties in the low-frequency regime, we obtain the results presented in Fig.~S3.

As shown in Fig.~S3(a) and (c), we establish two zones along the stacking $y$-direction. In both zones, the gradients of the $\alpha$ and $\beta$ chains are identical; the distinction lies in the coupling strength $t_\perp$. Specifically, $t_\perp > t_c$ in the upper zone, while $t_\perp < t_c$ in the lower zone. These differing $t_\perp$ strengths in the upper and lower zones compete with one another. When the $t_\perp$ value of a specific zone dominates, the global skin mode aligns with the distribution characteristic of that zone (a, c). Analogous to the 1D results, the global configuration of the 2D system is governed by the winding direction of the low-frequency bands around the energy point $E_0$ in the complex plane (indicated by the blue curves in b and d). Even within this partitioned 2D structure, the winding direction near the low-frequency $E_0$ remains unidirectional (as shown by the blue curves). However, the competition between the $t_\perp$ parameters of the zones alters the winding direction around $E_0$, thereby shifting the distribution of the global skin mode observed in Fig.~S3(a,c).

\section{IV. NONLINEAR EXTENSION AND AMPLITUDE-DRIVEN TOPOLOGICAL TRANSITION}

\begin{figure*}[htpb]
\renewcommand{\thefigure}{S4}
\includegraphics[width=\linewidth]{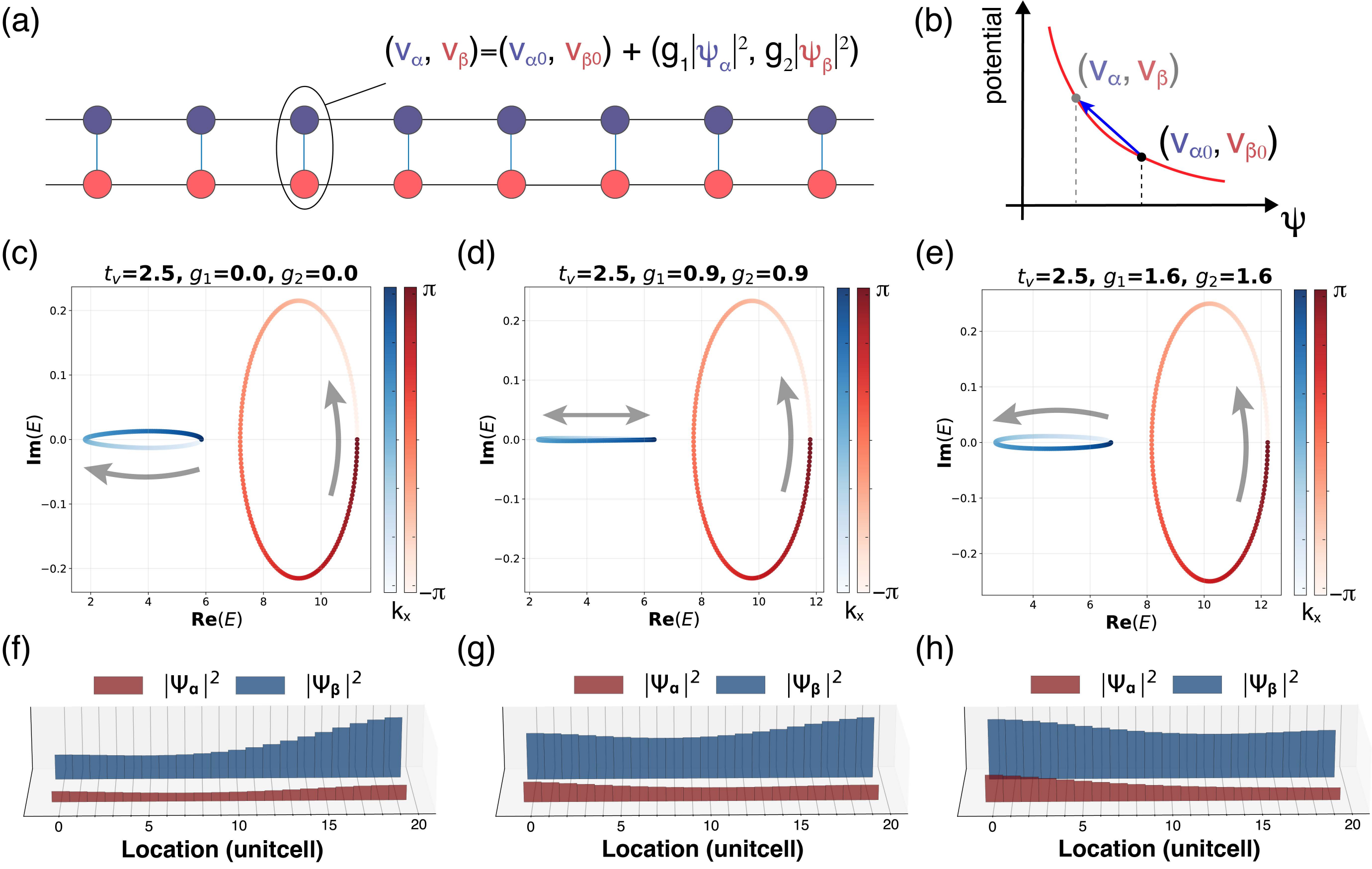}
\caption{\label{fig:S4} (a) Schematic of the nonlinear double-chain model where the system-reservoir coupling is amplitude-dependent. (b) Physical realization using magnetic springs, which provide a nonlinear force-displacement response (red curve) acting as the Kerr-like interaction. (c)--(e) Evolution of the complex energy spectrum in the complex plane for increasing nonlinear strength $g_1=g_2$: (c) $g=0$ (linear regime), (d) $g=0.9$ (critical point), and (e) $g=1.6$ (strongly nonlinear). The spectral winding direction (gray arrows) inverts, indicating a topological transition driven purely by nonlinearity. (f)--(h) Corresponding spatial profiles of the vibration amplitude. As the nonlinearity increases, the skin mode localization reverses from the right boundary to the left boundary, confirming the amplitude-dependent renormalization of the effective decay gradients.}
\end{figure*}

The paradigm of structured reservoir engineering established in Section I implies that the effective non-Hermiticity is dictated by the system-reservoir interaction profile $v(x)$. If the interaction elements exhibit nonlinearity, the self-energy $\Sigma$ becomes dependent on the eigenmode amplitude $|\psi|^2$, leading to intrinsic topological phase transitions driven by signal intensity. To investigate this, we extend the double-chain model by introducing Kerr-like nonlinearity into the grounded springs. Physically, this can be realized by replacing the linear springs with magnetic-like interactions, as illustrated in Fig.~S4(a,b). The restoring force becomes nonlinear, $F_{j} \approx k_j u_j + \alpha_j u_j^2$, implying that the effective stiffness (and thus the coupling strength to the reservoir) depends on the local displacement amplitude. Consequently, the effective Hamiltonian takes the form $H_{\text{eff}}(\psi) = H_{SS} + \Sigma(\omega, |\psi|^2)$.

We focus on the case where the nonlinearity modifies the variation of the effective decay potentials $\Gamma_\alpha(x, |\psi|)$ and $\Gamma_\beta(x, |\psi|)$. Specifically, we model the onsite potentials as $V_{j} = V_{j,0} + g_j |\psi_j|^2$, where $g_j$ quantifies the nonlinear strength. Using a self-consistent iterative algorithm, we solved for the eigenstates of the nonlinear system.

The results, presented in Fig.~S4(c-e), reveal a profound amplitude-driven phenomenon. With the passive inter-chain coupling $t_\perp<t_c$ fixed, increasing the nonlinear strength $g$ (or equivalently, the excitation amplitude) effectively reshapes the competition between the decay gradients of the $\alpha$ and $\beta$ chains. At a critical intensity, the dominant decay channel switches, inverting the sign of the effective imaginary gauge field. This triggers a topological phase transition where the winding number flips from $w=1$ to $w=-1$, accompanied by a reversal of the skin mode localization [Fig.~S4(f-h)]. This finding confirms that nonlinearity can serve as an intrinsic control dimension to dynamically manipulate the non-Hermitian topology in passive systems.

\section{V. SENSITIVITY AND TUNABILITY ANALYSIS}

\begin{figure}[htpb]
    \centerline{\includegraphics[width=0.6\linewidth]{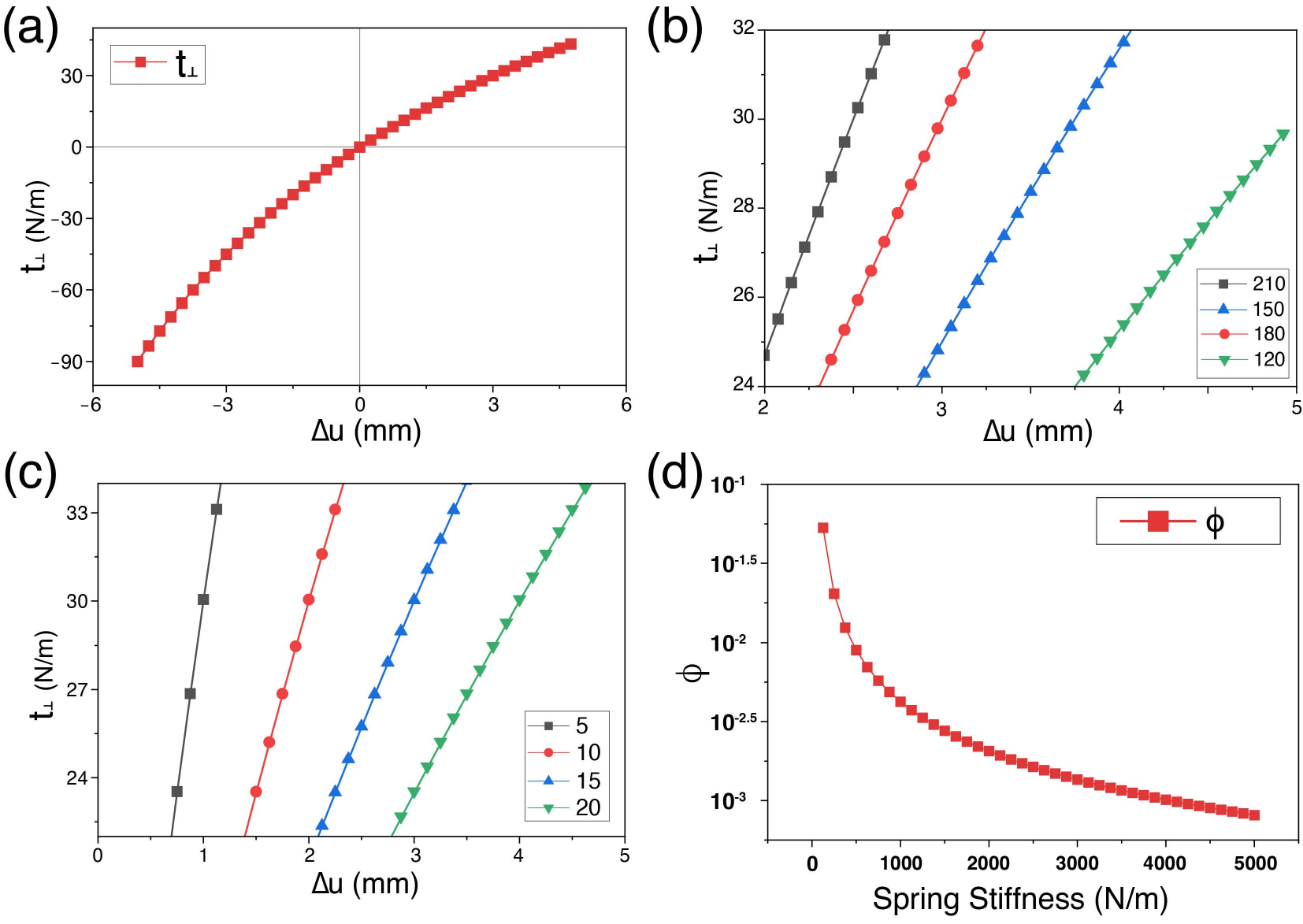}}
    \begin{flushleft}\label{fig:S3}FIG. S5. Tunability and sensitivity analysis of the mechanical prototype. (a) Calibration curve of the effective inter-chain coupling $t_\perp$ as a function of the spring stretching length $\Delta u$, for a spring with stiffness $k=\SI{179.13}{N/m}$. (b, c) Sensitivity of $t_\perp$ to length changes for different spring stiffnesses (b) and equilibrium lengths (c). (d) Calculated dimensionless precision $\phi = \Delta l/l$ required to trigger the topological transition as a function of spring stiffness. Using stiffer springs can significantly enhance sensitivity, potentially reaching $\phi \sim 10^{-6}$ for stiffness on the order of $10^6$ \si{N/m}. \end{flushleft}
\end{figure}

The critical point $t_c$ for the topological phase transition is determined by the competition between the intra-chain hopping $t$ and the reservoir-induced decay gradients. Once the system is designed near this critical point, the skin mode localization becomes extremely sensitive to parameter variations.

We characterize this sensitivity by the fractional change in the spring length $\phi = \Delta l / l$ required to trigger the reversal. As shown in Fig.~S5(a), the effective coupling $t_\perp$ exhibits a nonlinear dependence on the spring stretching length $\Delta u$. In Fig. S5(b), we illustrate the effect of stiffness $k$ on $t_\perp$'s variation. Fig. S5(c) shows the impact of different connection lengths $l_0+\Delta u$ on $t_\perp$. From these observations, we can conclude that a greater stiffness $k$ and shorter length $l_0+\Delta u$ result in the system responding to smaller changes in $\Delta u$ for phase transition, thereby exhibiting higher sensitivity. Fig.~S5(d) plots the calculated precision $\phi$ as a function of spring stiffness. With our current setup ($k \approx \SI{179}{N/m}$), the transition is triggered by a length change of $\approx \SI{3}{mm}$, corresponding to $\phi \approx 0.02$. Theoretical extrapolation indicates that using stiffer springs (on the order of $10^3$--$10^6$ \si{N/m}) could enhance the sensitivity to $\phi \sim 10^{-6}$, suggesting the potential of this passive platform for high-precision sensing applications without the need for active gain stabilization.

\section{VI. EXTENSIONS TO MICROSCALE AND ELECTRICAL SYSTEMS}

\begin{figure*}[htpb]
\begin{center}
\centerline{\includegraphics[width=0.6\linewidth]{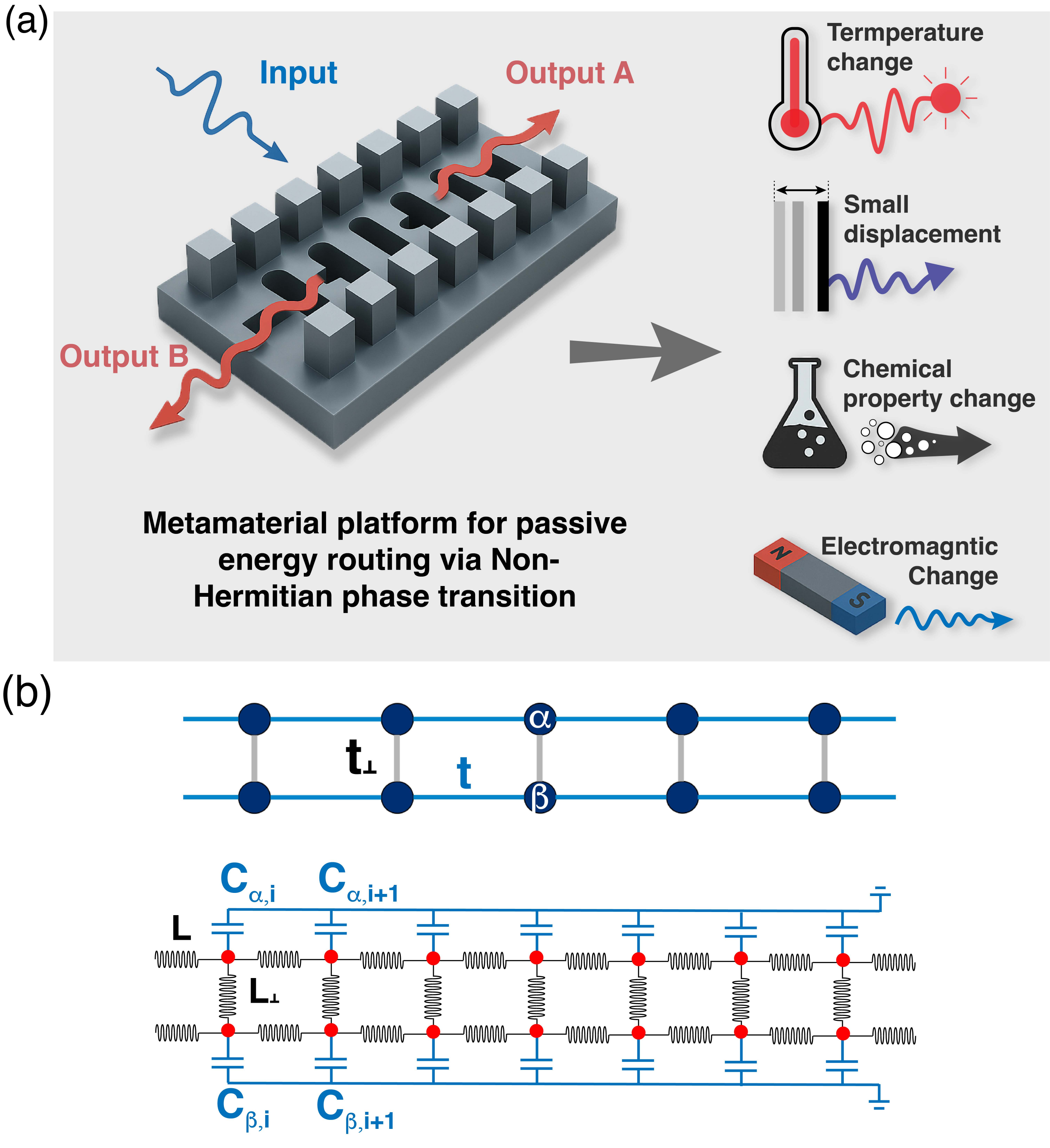}}
\begin{flushleft}{\label{fig.s5} FIG. S6 (a) Conceptual extensions of the passive non-Hermitian paradigm to other physical domains. (a) Schematic of a MEMS/NEMS implementation for adaptive energy routing. The manual tuning of $t_\perp$ is replaced by environmental stimuli (e.g., temperature, chemical binding) that intrinsically modulate the coupling stiffness, driving the system across the topological transition point $t_c$. (b) Equivalent circuit analog. The mechanical equations of motion map to the telegrapher's equations for an LC network, where the structured reservoir is realized by a graded array of grounding capacitors. This isomorphism indicates that the passive NHSE can be implemented in purely passive electrical circuits.} \end{flushleft}
\end{center}
\end{figure*}

While our experimental validation is macroscopic, the established paradigm of structured reservoir engineering is scale-invariant and can be generalized to other wave domains.
The principles demonstrated here are directly transferable to Micro/Nano-Electro-Mechanical Systems (MEMS/NEMS). In such platforms, the manually tuned spring $t_\perp$ can be replaced by coupling elements sensitive to environmental stimuli, such as thermal expansion beams or chemically responsive polymers (see Fig.~S6). A local environmental change (e.g., temperature crossing a threshold $T_c$) would intrinsically modulate the effective stiffness, driving the system across the topological transition and passively redirecting vibrational energy from one port to another.

The mechanical equations of motion map exactly to the telegrapher's equations for an LC circuit network. As illustrated in Fig.~S6(b), the displacement $u_i$ maps to the voltage node flux $\psi_i = \int V_i dt$, the mass $m$ to capacitance $C$, and the spring stiffness $k$ to inverse inductance $L^{-1}$. The structured reservoir corresponds to a graded array of capacitors grounding the transmission line. The equation of motion for the circuit analog is:
\begin{equation}
C_{\alpha,i} \ddot{\psi}_{\alpha,i} = -(2L^{-1} + L_\perp^{-1})\psi_{\alpha,i} + L^{-1}(\psi_{\alpha,i+1} + \psi_{\alpha,i-1}) + L_\perp^{-1}\psi_{\beta,i},
\end{equation}
which is mathematically isomorphic to Eq.~(1) in the main text. This isomorphism implies that the passive non-Hermitian skin effect and its topological reversal can be implemented in purely passive electrical circuits for robust signal routing and sensing applications.
